\documentclass[submission,copyright,creativecommons]{eptcs}
\providecommand{\event}{WoF 2015} 
\usepackage{breakurl}             
\usepackage{latexsym}
\usepackage{lingmacros,xypic,prooftree,amssymb,lscape}

\newcommand{\techterm}[1]{\mbox{\it #1}} 
\newcommand{\disp}[1]{\enumsentence{#1}}

\newcommand{\tb}{\hspace*{0.25in}}
\newcommand{\tab}{\hspace*{0.5in}}

\newcommand{\oscott}{[[}
\newcommand{\cscott}{]]}
\newcommand{\vtab}{\ \vspace{2.5ex}\\}

\newcommand{\yields}{\mbox{\ $\Rightarrow$\ }}

\newcommand{\add}{\mbox{$+$}}

\newcommand{\wrap}{\mbox{$\times$}}
\newcommand{\bsl}{\mbox{$\backslash$}}
\newcommand{\product}{\mbox{$\bullet$}}
\newcommand{\infix}{\mbox{$\downarrow$}}

\newcommand{\circum}{\mbox{$\uparrow$}}
\newcommand{\wprod}{\mbox{$\odot$}}
\newcommand{\sinfix}[1]{\mbox{$\downarrow_{#1}$}}
\newcommand{\swrap}[1]{\mbox{$\wrap{_{#1}}$}}
\newcommand{\smwrap}[1]{\mbox{$\,|{_{#1}}\,$}}
\newcommand{\mwrap}[1]{\mbox{$|$}}
\newcommand{\scircum}[1]{\mbox{$\uparrow_{#1}$}}
\newcommand{\swprod}[1]{\mbox{$\odot{_{#1}}$}}
\newcommand{\D}{\mbox{\bf D}}

\newcommand{\prf}[1]{\noindent{\bf Proof}. #1 $\square$}
\newcommand{\abrack}{\mbox{$[\,]^{-1}$}}
\newcommand{\mybrack}{\mbox{$\langle\rangle$}}
\newcommand{\sep}{\mbox{$1$}}
\newcommand{\vect}[1]{\overrightarrow{#1}}
\newcommand{\aconj}{\mbox{$\&$}}
\newcommand{\adisj}{\mbox{$\oplus$}}

\newcommand{\inp}{\mbox{$^{\bullet}$}}
\newcommand{\out}{\mbox{$^{\circ}$}}

\newcommand{\ass}{\mbox{$:\,$}}
\newcommand{\sass}{\mbox{$;\,$}}

\newcommand{\subst}[1]{\mbox{$\{#1\}$}}
\newcommand{\zero}{\mbox{$0$}}
\newcommand{\commentout}[1]{}

\newcommand{\syncnst}[1]{\mbox{\it #1}}
\newcommand{\CN}{\mbox{$\mathit{CN}$}}
\newcommand{\QP}{\mbox{$\mathit{QP}$}}
\newcommand{\VP}{\mbox{$\mathit{VP}$}}
\newcommand{\TV}{\mbox{$\mathit{TV}$}}

\newtheorem{theorem}{Theorem}[section]

\title{Multiplicative-Additive Focusing for Parsing as Deduction}
\author{Glyn Morrill \qquad\qquad
Oriol Valent\'{\i}n
\institute{Department of Computer Science\\
Universitat Polit\`ecnica de Catalunya\\
Barcelona}
 \email{morrill@cs.upc.edu} \qquad\qquad \email{oriol.valentin@gmail.com}}

\def\titlerunning{MA focusing for Parsing}
\def\authorrunning{Morrill and Valent\'{\i}n}

\newcommand{\mini}{}

\newcommand{\AL}{\mbox{\bf L}}
\newcommand{\DA}{\mbox{\bf DA}}
\newcommand{\DAf}{\mbox{$\mathbf{DA_{foc}}$}} 
\newcommand{\lyieldsw}{\mbox{$\Longrightarrow_w$}} 
\newcommand{\lyieldsws}{\mbox{$\Longrightarrow$}} 
\newcommand{\DAF}{\mbox{$\mathbf{DA_{Foc}}$}}
\newcommand{\lyields}{\mbox{$\Longrightarrow$}} 
\newcommand{\pf}{\mbox{ $\Diamond$ foc}}
\newcommand{\foc}{\mbox{\it foc}} 
\newcommand{\pcut}{\mbox{\it{p-Cut}}} 
\newcommand{\ncut}{\mbox{\it{n-Cut}}} 

\spreaddiagramrows{-1pc}
\spreaddiagramcolumns{-1pc}

\begin{document}

\maketitle

\begin{abstract}
Spurious ambiguity is the phenomenon whereby distinct derivations in grammar
may assign the same structural reading,
resulting in redundancy in the parse search space and inefficiency in parsing.
Understanding the problem depends on identifying the essential mathematical structure
of derivations.
This is trivial in the case of context free grammar, where the parse structures
are ordered trees;
in the case of type logical categorial grammar, the parse structures
are proof nets. However, with respect to multiplicatives intrinsic 
proof nets have not yet been given for displacement calculus, and proof nets for additives,
which have applications to polymorphism, are not easy to characterise. 
Here we approach multiplicative-additive spurious ambiguity
by means of the proof-theoretic technique of focalisation.
\end{abstract}

\section{Introduction}

In context free grammar (CFG) sequential rewriting derivations exhibit spurious ambiguity:
distinct rewriting derivations may correspond to the same parse structure (tree) and the
same structural reading.\footnote{Research partially supported by
SGR2014-890 (MACDA) of the Generalitat de Catalunya and 
MINECO project APCOM (TIN2014-57226-P), and by
an ICREA Acad\`emia 2012 to GM.
Thanks to three anonymous WoF reviewers for comments and suggestions,
and to Iliano Cervesato for editorial attention. All errors are our own.}
In this case it is transparent to develop parsing algorithms avoiding spurious ambiguity
by reference to parse trees. 
In categorial grammar (CG) the problem is more subtle. The Cut-free Lambek
sequent proof search space is finite, but involves a combinatorial explosion
of spuriously ambiguous sequential proofs.
This can be understood, analogously to CFG, as inessential
rule reorderings, which we parallelise in underlying geometric parse structures
which are (planar) proof nets.

The planarity of Lambek proof nets reflects that the formalism is continuous or
concatenative. But the challenge of natural grammar is discontinuity or
apparent displacement, whereby there is syntactic/semantic mismatch,
or elements appearing out of place. Hence the subsumption of Lambek
calculus by displacement calculus \D{} including intercalation as well as
concatenation \cite{mvf:tdc}.

Proof nets for \D{} must be partially non-planar; steps
towards intrinsic correctness criteria for displacement proof nets are
made in 
\cite{fadda:phd}
and 
\cite{moot:lamfes}. 
Additive proof nets are considered in \cite{hughesglabbeck:05} and \cite{abruscimaieli:fg15}.
However, even in the case
of Lambek calculus, parsing by reference to intrinsic criteria 
\cite{mootret},
\cite{morrill:oxford}, appendix B, 
is not more efficient than parsing by reference to
extrinsic criteria of normalised sequent calculus 
\cite{hendriksH:phd}.
In its turn, on the other hand,
normalisation does not extend to product left rules and product unit
left rules nor to additives.
The focalisation of 
\cite{andreoli:92}
is a methodology midway between proof nets and normalisation.
Here we apply the focusing discipline 
to the parsing as deduction of \D{}
with additives.

In \cite{cms:08} multifocusing is defined for unit-free MALL,\footnote{Here we include units,
which are linguistically relevant.} providing canonical sequent
proofs; an eventual goal
would be to formulate multifocusing for multiplicative-additive categorial logic
and for categorial logic generally. In this respect the present paper represents
an intermediate step. Note that \cite{simmons:12} develops focusing for
Lambek calculus with additives, but not for displacement logic, for which
we show completeness of focusing here.

\commentout{
The paper is as follows.
In Sections~\ref{spambcfg} and~\ref{spamblc} we describe spurious ambiguity
in context-free grammar and Lambek calculus.
In Section~\ref{tdc} we recall the displacement calculus with additives.
In Section~\ref{foc} we present focalisation for the displacement calculus with additives.
In Section~\ref{compl} we prove completeness of the focalisation for displacement calculus and 
we conclude in Section~\ref{concl} with an example.
}

\subsection{Spurious ambiguity in CFG}

\label{spambcfg}

Consider the following production rules:
$$\mini
\begin{array}{l}
S \rightarrow \QP\ \VP\\
\QP \rightarrow Q\ \CN\\
\VP \rightarrow \TV\ N
\end{array}
$$
These generate the following sequential rewriting derivations:
$$\mini
\begin{array}{l}
S \rightarrow \QP\ \VP \rightarrow Q\ \CN\ \VP \rightarrow Q\ \CN\ \TV\ N\\
S \rightarrow \QP\ \VP \rightarrow \QP\ \TV\ N \rightarrow Q\ \CN\ \TV\ N
\end{array}
$$
These sequential rewriting derivations correspond to the
same parellelised parse structure:
$$\mini
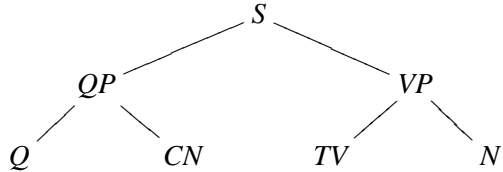
\diagram
 &&& S\\
 &\QP\urrline &&&&\VP\ullline\\
Q\urline && \CN\ulline && \TV\urline && N\ulline
\enddiagram
$$
And they correspond to the same structural reading;
sequential rewriting has \techterm{spurious ambiguity}.

\subsection{Spurious ambiguity in CG}

\label{spamblc}

Lambek calculus is a logic of strings with the operation $\add$ of concatenation.
Recall the definitions of types, configurations and sequents in the Lambek
calculus \AL{} 
\cite{lambek:mathematics}
, in terms of a set $\cal P$ of primitive types
(the original Lambek calculus did not include the product unit):
\disp{\mini
$\begin{array}[t]{lrcl}
\mbox{Types} & \cal F & ::= & {\cal P}\ |\ {\cal F}/{\cal F}\ |\ {\cal F}\bsl{\cal F}\ |\ {\cal F}\product{\cal F}\\
\mbox{Configurations} & {\cal O} & ::= & \Lambda\ |\ {\cal F}, {\cal O}\\
\mbox{Sequents} & \Sigma & ::= & {\cal O}\yields{\cal F}
\end{array}$}
Lambek calculus types have the following interpretation:
$$
\begin{array}{rcl}
\oscott C/B\cscott & = & \{s_1|\ \forall s_2\in\oscott B\cscott, s_1\add s_2\in\oscott C\cscott\}\\
\oscott A\bsl C\cscott & = & \{s_2|\ \forall s_1\in\oscott A\cscott, s_1\add s_2\in\oscott C\cscott\}\\
\oscott A\product B\cscott & = & \{s_1\add s_2|\ s_1\in\oscott A\cscott\ \&\ s_2\in\oscott B\cscott\}
\end{array}
$$
The logical rules of \AL{} are as follows:

\begin{center}
\mini$
\prooftree
\Gamma\yields A \tb
\Delta(C)\yields D
\justifies
\Delta(\Gamma, A\bsl C)\yields D
\using \bsl L
\endprooftree \tb
\prooftree
A, \Gamma\yields C
\justifies
\Gamma\yields A\bsl C
\using \bsl R
\endprooftree$
\vtab
$\prooftree
\Gamma\yields B\tb
\Delta(C)\yields D
\justifies
\Delta(C/B,  \Gamma)\yields D
\using / L
\endprooftree \tb
\prooftree
\Gamma, B\yields C
\justifies
\Gamma\yields C/B
\using /R
\endprooftree$
\vtab
$\prooftree
\Delta(A, B)\yields D
\justifies
\Delta(A\product B)\yields D
\using \product L
\endprooftree$
\tb
$\prooftree
\Gamma_1\yields A\tb\Gamma_2\yields B
\justifies
\Gamma_1, \Gamma_2\yields A\product B
\using \product R
\endprooftree
$\\
\end{center}

\begin{figure*}
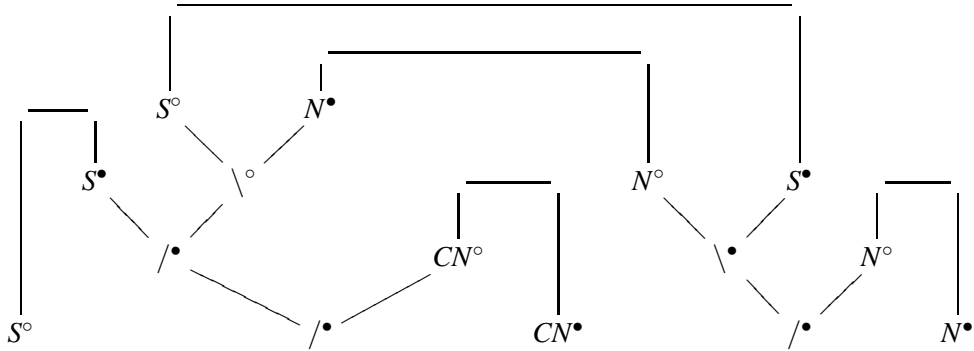

\begin{center}\mini
$
\diagram
&&&&&&&&&&&&\\
&&&&&&&&&&&&\\
&& S\out && N\inp\\
& S\inp && \bsl\out\ulline\urline &&&&& N\out\xline'[-2,0]'[-2,-4]'[-1,-4] && S\inp\xline'[-3,0]'[-3,-8]'[-1,-8]&&\\
&& /\inp\ulline\urline&&&& \CN\out&&&\bsl\inp\ulline\urline&& N\out\\
S\out\xline'[-3,0]'[-3,1]'[-2,1]&&&&/\inp\ullline\urrline&&&\CN\inp\xline'[-2,0]'[-2,-1]'[-1,-1]&&&/\inp\ulline\urline && N\inp\xline'[-2,0]'[-2,-1]'[-1,-1]
\enddiagram
$
\end{center}
\caption{Proof net}
\label{pn}
\end{figure*}

\noindent
Even amongst Cut-free proofs there is spurious ambiguity; consider for example
the sequential derivations of Figure~\ref{seqder}.
\begin{figure*}\small
\begin{center}
\prooftree
\CN\yields\CN
\prooftree
\prooftree
\prooftree
N\yields N
\prooftree
N\yields N\tb S\yields S
\justifies
N, N\bsl S\yields S
\using \bsl L
\endprooftree
\justifies
N, (N\bsl S)/N, N\yields S
\using /L
\endprooftree
\justifies
(N\bsl S)/N, N\yields N\bsl S
\using \bsl R
\endprooftree
S\yields S
\justifies
S/(N\bsl S), (N\bsl S)/N, N\yields S
\using /L
\endprooftree
\justifies
(S/(N\bsl S))/\CN, \CN, (N\bsl S)/N, N\yields S
\using  /L
\endprooftree
\tb
\prooftree
N\yields N
\prooftree
\CN\yields\CN
\prooftree
\prooftree
\prooftree
N\yields N\tb S\yields S
\justifies
N, N\bsl S\yields S
\using \bsl L
\endprooftree
\justifies
N\bsl S\yields N\bsl S
\using \bsl R
\endprooftree
S\yields S
\justifies
S/(N\bsl S), N\bsl S\yields S
\using /L
\endprooftree
\justifies
(S/(N\bsl S))/\CN, \CN, N\bsl S\yields S
\using /L
\endprooftree
\justifies
(S/(N\bsl S))/\CN, \CN, (N\bsl S)/N, N\yields S
\using /L
\endprooftree
\end{center}
\caption{Spurious ambiguity}
\label{seqder}
\end{figure*}
These have the same parallelised parse structure (proof net) of Figure~\ref{pn}.

Lambek proof structures are planar
graphs which must satisfy certain global and local properties to be correct as proofs
(proof nets). Proof nets provide a geometric perspective
on derivational equivalence. Alternatively we may identify the same
algebraic parse structure (Curry-Howard term):
$$((x_{\mathit{Q}}\,x_{\mathit{CN}})\,\lambda x((x_{\mathit{TV}}\,x_{N})\,x))$$
But Lambek calculus is continuous (planarity).
A major issue issue in grammar is discontinuity, hence 
the displacement calculus.

\section{D with additives, DA}

\label{tdc}

\begin{figure}
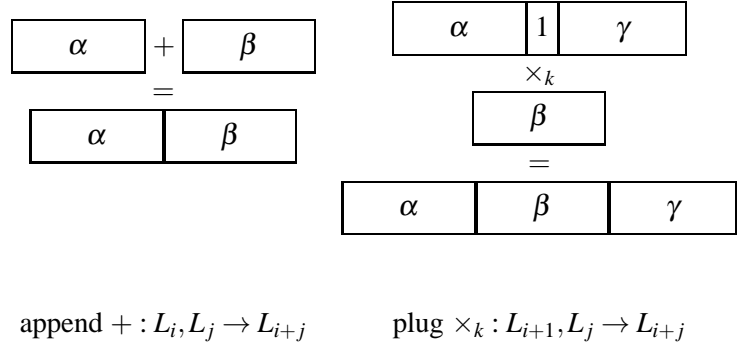

$$
\begin{array}[b]{c}
\fbox{$\tb\alpha\tb$\strut}\ \add\ \fbox{$\tb\beta\tb$\strut}\\
=\\
\fbox{$\tb\alpha\tb$\strut}\fbox{$\tb\beta\tb$\strut}\\\\\\\\\\
\mbox{append\ } +: L_i, L_j\rightarrow L_{i{+}j}
\end{array}
\begin{array}[b]{c}
\fbox{$\tb\alpha\tb$\strut}\fbox{$1$\strut}\fbox{$\tb\gamma\tb$\strut}\\
\swrap{k}\\
\fbox{$\tb\beta\tb$\strut}\\
=\\
\fbox{$\tb\alpha\tb$\strut}\fbox{$\tb\beta\tb$\strut}\fbox{$\tb\gamma\tb$\strut}\\\\\\
\mbox{plug\ } \times_k: L_{i{+}1}, L_j\rightarrow L_{i{+}j}
\end{array}
$$
\caption{Append and plug}
\label{diag}
\end{figure}

In this section we present displacement calculus {\bf D},
and a displacement logic {\bf DA} comprising {\bf D} with additives.
Although {\bf D} is indeed a conservative extension of {\bf L},
we think of it not just as an \emph{extension\/} of Lambek
calculus but as a \emph{generalisation}, because
it involves a whole new machinery of sequent calculus to
deal with discontinuity.
Displacement calculus  is a logic of discontinuous strings --- strings punctuated
by a \techterm{separator} $\sep$ and subject to operations of append and plug;
see Figure~\ref{diag}.
Recall the definition of types and their sorts, 
configurations and their sorts,
and sequents, for the displacement calculus 
with additives:

\disp{\mini
Types 
\begin{array}[t]{rclrclll}
{\cal F}_i & ::= & {\cal F}_{i{+}j}/{\cal F}_j\\
{\cal F}_j & ::= & {\cal F}_i\bsl{\cal F}_{i{+}j}\\
{\cal F}_{i{+}j} & ::= & {\cal F}_i\product{\cal F}_j\\
{\cal F}_0 & ::= & I\\
{\cal F}_{i{+}1} & ::= & {\cal F}_{i{+}j}\scircum{k}{\cal F}_j& 1\le k\le i{+}1\\
{\cal F}_j & ::= & {\cal F}_{i{+}1}\sinfix{k}{\cal F}_{i{+}j}& 1\le k\le i{+}1\\
{\cal F}_{i{+}j} & ::= & {\cal F}_{i{+}1}\swprod{k}{\cal F}_j& 1\le k\le i{+}1\\
{\cal F}_1 & ::= & J\\
{\cal F}_i & ::= & {\cal F}_{i}\aconj{\cal F}_i\\
{\cal F}_i & ::= & {\cal F}_{i}\adisj{\cal F}_i\\
\end{array}\\\\
Sort $sA = \mbox{the } i \mathrm{\ s.t.\ }A\in {\cal F}_i$\\
For example, $s(S\scircum{1}N)\scircum{2}N=s(S\scircum{1}N)\scircum{1}N=2$
where $sN=sS=0$\\\\
Configurations $\begin{array}[t]{rcl}{\cal O} & ::= &
 \Lambda\ |\ {\cal T}, {\cal O}\\
 {\cal T} & ::= & \sep\ |\ {\cal F}_0\ |\ {\cal F}_{i{>}0}\{\underbrace{{\cal O}:
\ldots : {\cal O}}_{i\,{\cal O}\mathrm{'s}}\}
\end{array}$\\
For example, there is the configuration
$(S\scircum{1}N)\scircum{2}N\{N, \sep: S\scircum{1}N, S\}, \sep, N, \sep$\\\\
Sort $s{\cal O} = |{\cal O}|_{\mathrm{1}}$\\
For example $s(S\scircum{1}N)\scircum{2}N\{N, \sep: S\scircum{1}N, S\}, \sep, N, \sep=3$\\\\
Sequents $\Sigma ::= {\cal O}\yields A\mathrm{\ s.t.\ }s{\cal O} = sA$
}

The figure $\vect{A}$ of a type $A$ is defined by:
$$\mini
\vect{A} = \left\{
\begin{array}{ll}
A & \mbox{if\ } sA=0\\
A\{\underbrace{\sep: \ldots: \sep}_{sA\ 1\mathit{'s}}\} & \mbox{if\ } sA>0
\end{array}\right.$$

Where $\Gamma$ is a configuration of sort $i$ and $\Delta_1, \ldots, \Delta_i$
are configurations,
the \techterm{fold} $\Gamma\otimes\langle\Delta_1: \ldots: \Delta_i\rangle$
is the result of replacing the successive \sep{}'s in $\Gamma$
by $\Delta_1, \ldots, \Delta_i$ respectively.


Where $\Delta$ is a configuration of sort $i>0$ and $\Gamma$ is a configuration, the
$k$\/th metalinguistic wrap $\Delta\smwrap{k}\Gamma$, $1\le k\le i$,  is given by
\disp{\mini$
\Delta\smwrap{k}\Gamma =_{df} \Delta\otimes\langle\underbrace{\sep: \ldots: \sep}_{k{-}1\ 1\mbox{\footnotesize 's}}: \Gamma: \underbrace{\sep: \ldots: \sep}_{i{-}k\ 1\mbox{\footnotesize 's}}\rangle$
}
\noindent
i.e.\ $\Delta\smwrap{k}\Gamma$ is the configuration
resulting from replacing by $\Gamma$ the $k$\/th separator
in $\Delta$.

In broad terms, syntactical interpretation of displacement calculus is as follows:
$$
\begin{array}{rcl}
\oscott C/B\cscott & = & \{s_1|\ \forall s_2\in\oscott B\cscott, s_1\add s_2\in\oscott C\cscott\}\\
\oscott A\bsl C\cscott & = & \{s_2|\ \forall s_1\in\oscott A\cscott, s_1\add s_2\in\oscott C\cscott\}\\
\oscott A\product B\cscott & = & \{s_1\add s_2|\ s_1\in\oscott A\cscott\ \&\ s_2\in\oscott B\cscott\}\\
\oscott I\cscott & = & \{0\}
\\\\
\oscott C\scircum{k}B\cscott & = & \{s_1|\ \forall s_2\in\oscott B\cscott, s_1\swrap{k} s_2\in\oscott C\cscott\}\\
\oscott A\sinfix{k} C\cscott & = & \{s_2|\ \forall s_1\in\oscott A\cscott, s_1\swrap{k} s_2\in\oscott C\cscott\}\\
\oscott A\swprod{k} B\cscott & = & \{s_1\swrap{k} s_2|\ s_1\in\oscott A\cscott\ \&\ s_2\in\oscott B\cscott\}\\
\oscott J\cscott & = & \{1\}
\end{array}
$$

The logical rules of the displacement calculus with additives are as follows,
where
$\Delta\langle\Gamma\rangle$
abbreviates $\Delta_0(\Gamma\otimes\langle\Delta_1: \ldots: \Delta_i\rangle)$:\\

\begin{center}\mini
$\prooftree
\Gamma\yields B \tb
\Delta\langle\vect{C}\rangle\yields D
\justifies
\Delta\langle\vect{C/B}, \Gamma\rangle\yields D
\using / L
\endprooftree \tb
\prooftree
\Gamma, \vect{B}\yields C
\justifies
\Gamma\yields C/B
\using / R
\endprooftree$
\vtab
$\prooftree
\Gamma\yields A \tb
\Delta\langle\vect{C}\rangle\yields D
\justifies
\Delta\langle\Gamma, \vect{A\bsl C}\rangle\yields D
\using \bsl L
\endprooftree \tb
\prooftree
\vect{A}, \Gamma\yields C
\justifies
\Gamma\yields A\bsl C
\using \bsl R
\endprooftree$
\vtab
$\prooftree
\Delta\langle\vect{A}, \vect{B}\rangle\yields D
\justifies
\Delta\langle\vect{A\product B}\rangle\yields D
\using \product L
\endprooftree
\tb
\prooftree
\Gamma_1\yields A\tb\Gamma_2\yields B
\justifies
\Gamma_1, \Gamma_2\yields A\product B
\using \product R
\endprooftree
$
\vtab
\prooftree
\Delta\langle\Lambda\rangle\yields A
\justifies
\Delta\langle\vect{I}\rangle\yields A
\using IL
\endprooftree\tb
\prooftree
\justifies
\Lambda\yields I
\using IR
\endprooftree
\end{center}\ \\

\begin{center}\mini
$\prooftree
\Gamma\yields B \tb
\Delta\langle\vect{C}\rangle\yields D
\justifies
\Delta\langle\vect{C\scircum{k} B}\smwrap{k}\Gamma\rangle\yields D
\using \scircum{k} L
\endprooftree \tb
\prooftree
\Gamma\smwrap{k}\vect{B}\yields C
\justifies
\Gamma\yields C\scircum{k} B
\using \scircum{k} R
\endprooftree$
\vtab
$\prooftree
\Delta\langle\vect{A}\smwrap{k}\vect{B}\rangle\yields D
\justifies
\Delta\langle\vect{A\swprod{k} B}\rangle\yields D
\using \swprod{k} L
\endprooftree
\tb
\prooftree
\Gamma_1\yields A\tb\Gamma_2\yields B
\justifies
\Gamma_1\smwrap{k}\Gamma_2\yields A\swprod{k} B
\using \swprod{k} R
\endprooftree$
\vtab
$\prooftree
\Gamma\yields A\tb
\Delta\langle\vect{C}\rangle\yields D
\justifies
\Delta\langle\Gamma\smwrap{k}\vect{A\sinfix{k} C}\rangle\yields D
\using \sinfix{k} L
\endprooftree \tb
\prooftree
\vect{A}\smwrap{k}\Gamma\yields C
\justifies
\Gamma\yields A\sinfix{k} C
\using \sinfix{k} R
\endprooftree$
\vtab
\prooftree
\Delta\langle\sep\rangle\yields A
\justifies
\Delta\langle\vect{J}\rangle\yields A
\using JL
\endprooftree\tb
\prooftree
\justifies
\sep\yields J
\using JR
\endprooftree
\vtab
\prooftree
\Gamma\langle\vect{A}\rangle\yields C
\justifies
\Gamma\langle\vect{A\aconj B}\rangle\yields C
\using \aconj L_1
\endprooftree\tb
\prooftree
\Gamma\langle\vect{B}\rangle\yields C
\justifies
\Gamma\langle\vect{A\aconj B}\rangle\yields C
\using \aconj L_2
\endprooftree
\vtab
\prooftree
\Gamma\yields A\tb\Gamma\yields B
\justifies
\Gamma\yields A\aconj B
\using \aconj R
\endprooftree
\vtab
\prooftree
\Gamma\langle\vect{A}\rangle\yields C\tb\Gamma\langle\vect{B}\rangle\yields C
\justifies
\Gamma\langle\vect{A\adisj B}\rangle\yields C
\using \adisj L
\endprooftree
\vtab
\prooftree
\Gamma\yields A
\justifies
\Gamma\yields A\adisj B
\using \adisj R_1
\endprooftree\tb
\prooftree
\Gamma\yields B
\justifies
\Gamma \yields A\adisj B
\using \adisj R_2
\endprooftree
\end{center}

The continuous multiplicatives  \{$/$, $\bsl$, $\product$, $I$\}
of 
Lambek (1958\cite{lambek:mathematics};
1988\cite{lambek:88}),
are the basic means of categorial (sub)categorization.
The directional divisions over, $/$, and under, $\bsl$, are exemplified
by assignments such as $\syncnst{the}\ass N/\CN$ for
$\syncnst{the man}\ass N$ and $\syncnst{sings}\ass N\bsl S$ for
$\syncnst{John sings}\ass S$,
and $\syncnst{loves}\ass (N\bsl S)/N$ for $\syncnst{John loves Mary}\ass S$. 
Hence, for \syncnst{the man}:
$$\mini
\prooftree
\CN\yields\CN\tb N\yields N
\justifies
N/\CN, N\yields N
\using /L
\endprooftree$$
And for \syncnst{John sings} and \syncnst{John loves Mary}:
$$\mini
\prooftree
N\yields N\tb S\yields S
\justifies
N, N\bsl S\yields S
\using \bsl L
\endprooftree
\tb
\prooftree
N\yields N
\prooftree
N\yields N\tb S\yields S
\justifies
N, N\bsl S\yields S
\using \bsl L
\endprooftree
\justifies
N, (N\bsl S)/N, N\yields S
\using /L
\endprooftree
$$

The continuous product $\product$ is exemplified by a `small clause'
assignment such as $\syncnst{considers}\ass (N\bsl S)/$ $(N\product(\CN/\CN))$
for $\syncnst{John considers Mary socialist}\ass S$. 
$$\mini
\prooftree
\prooftree
N\yields N\tb 
\prooftree
\prooftree
\CN\yields\CN\tb\CN\yields\CN
\justifies
\CN/\CN, \CN\yields \CN
\using /L
\endprooftree
\justifies
\CN/\CN\yields \CN/\CN
\using /R
\endprooftree
\justifies
N, \CN/\CN\yields N\product(\CN/\CN)
\using \product R
\endprooftree
\prooftree
N\yields N\tb S\yields S
\justifies
N, N\bsl S\yields S
\using \bsl L
\endprooftree
\justifies
N, (N\bsl S)/(N\product(\CN/\CN)), N, \CN/\CN\yields S
\using /L
\endprooftree$$
Of course this use of product is not essential: we could just as well
have used  $((N\bsl S)/(\CN/\CN))/N$ since in general we have both
$A/(C\product B)\yields (A/B)/C$ (currying) and
$(A/B)/C\yields A/(C\product B)$ (uncurrying). 

The discontinuous multiplicatives \{$\scircum{}$, $\sinfix{}$, $\swprod{}$, $J$\},
the displacement connectives, of
Morrill and Valent\'{\i}n (2010\cite{mv:disp}),
Morrill et al.\  (2011\cite{mvf:tdc}),
are defined in relation to intercalation. 
When the value of the $k$ subscript is one it may be omitted,
i.e.\ it defaults to one. 
Circumfixation, or extraction, 
$\scircum{}$,
is exemplified by a discontinuous idiom assignment
$\syncnst{gives}\add\sep\add\syncnst{the}\add\syncnst{cold}\add\syncnst{shoulder}\ass$ 
$(N\bsl S)\scircum{}N$ for
$\syncnst{Mary gives the man the}$ $\syncnst{cold shoulder}\ass S$: 
$$\mini
\prooftree
\prooftree
\CN\yields\CN\tb N\yields N
\justifies
N/\CN, \CN\yields N
\using /L
\endprooftree
\prooftree
N\yields N\tb S\yields S
\justifies
N, N\bsl S\yields S
\using \bsl L
\endprooftree
\justifies
N, (N\bsl S)\circum{}N\{N/\CN, \CN\}\yields S
\using \circum{}L
\endprooftree$$ 

Infixation, 
$\sinfix{}$,
and extraction together are exemplified by a quantifier assignment
$\syncnst{everyone}\ass(S\scircum{} N)\sinfix{}S$
simulating Montague's S14 quantifying in: 
$$\mini
\prooftree
\prooftree
\ldots, N, \ldots\yields S
\justifies
\ldots, \sep, \ldots\yields S\circum{}N
\using \circum{}R
\endprooftree
\prooftree
\justifies
S\yields S
\using \mbox{\it id}
\endprooftree
\justifies
\ldots, (S\circum{} N)\infix{} S, \ldots\yields S
\using \infix{}L
\endprooftree$$

Circumfixation and discontinuous product,
$\swprod{}$,
are illustrated in an assignment to a relative pronoun
$\syncnst{that}\ass(\CN\bsl\CN)/((S\circum{}N)\swprod{}I)$
allowing both peripheral and medial extraction,
$\syncnst{that John likes}\ass\CN\bsl\CN$ and
$\syncnst{that John saw today}\ass\CN\bsl\CN$: 
$$\mini
\prooftree
\prooftree
\prooftree
N, (N\bsl S)/N, N\yields S
\justifies
N, (N\bsl S)/N, \sep\yields S\circum{}N
\using \circum{}R
\endprooftree
\prooftree
\justifies
\yields I
\using IL
\endprooftree
\justifies
N, (N\bsl S)/N\yields (S\circum{}N)\wprod I
\using \wprod R
\endprooftree
\CN\bsl\CN\yields \CN\bsl\CN
\justifies
(\CN\bsl\CN)/((S\circum N)\wprod I), N, (N\bsl S)/N\yields \CN\bsl\CN
\using /L
\endprooftree
$$

$$\mini
\prooftree
\prooftree
\prooftree
N, (N\bsl S)/N, N, S\bsl S\yields S
\justifies
N, (N\bsl S)/N, \sep, S\bsl S\yields S\circum{}N
\using \circum{}R
\endprooftree
\prooftree
\justifies
\yields I
\using IL
\endprooftree
\justifies
N, (N\bsl S)/N, S\bsl S\yields (S\circum{}N)\wprod I
\using \wprod R
\endprooftree
\CN\bsl\CN\yields \CN\bsl\CN
\justifies
(\CN\bsl\CN)/((S\circum N)\wprod I), N, (N\bsl S)/N, S\bsl S\yields \CN\bsl\CN
\using /L
\endprooftree
$$

The additive conjunction and disjunction \{$\aconj$, $\adisj$\}
of 
Lambek (1961\cite{lambek:61}), 
Morrill (1990\cite{morrill:galt90}), and Kanazawa (1992\cite{kanazawa:additives}),
capture polymorphism.
For example the additive conjunction $\aconj$ can be used for
$\syncnst{rice}\ass N\aconj\CN$ as in $\syncnst{rice grows}\ass S$
and $\syncnst{the rice grows}\ass S$:

\ \\

$$\mini
\prooftree
\prooftree
N\yields N
\justifies
N\aconj\CN\yields N
\using \aconj L_1
\endprooftree
S\yields S
\justifies
N\aconj\CN, N\bsl S\yields S
\using \bsl L
\endprooftree
\tb
\prooftree
N/\CN, \CN, N\bsl S\yields S
\justifies
N/\CN, N\aconj\CN, N\bsl S\yields S
\using \aconj L_2
\endprooftree
$$

\ \\

The additive disjunction $\adisj$ can be used for $\syncnst{is}\ass(N\bsl S)/(N\adisj(\CN/\CN))$
as in $\syncnst{Tully is Cicero}\ass S$ and $\syncnst{Tully is humanist}\ass S$:

\ \\

$$\mini
\prooftree
\prooftree
N\yields N
\justifies
N\yields N\adisj(\CN/\CN)
\using \adisj R_1
\endprooftree
N\bsl S\yields N\bsl S
\justifies
(N\bsl S)/(N\adisj(\CN/\CN)), N\yields N\bsl S
\using /L
\endprooftree
\tb
\prooftree
\prooftree
\CN/\CN\yields \CN/\CN
\justifies
\CN/\CN\yields N\adisj(\CN/\CN)
\using \adisj R_2
\endprooftree
N\bsl S\yields N\bsl S
\justifies
(N\bsl S)/(N\adisj(\CN/\CN)), \CN/\CN\yields N\bsl S
\using /L
\endprooftree
$$

{\section{Focalisation for DA}
\label{foc}

In focalisation situated (antecedent, input, \inp/ succedent, output, \out) non-atomic
types are classified as of negative (asynchronous)
or positive (synchronous) \emph{polarity\/} according as their rule is reversible
or not;
situated atoms are positive or negative according to their bias.
The table below summarizes the notational convention on formulas $P,Q,M$ and $N$:

\begin{center}$
  \begin{tabular}{| l | c |  c|}
    \hline
     & \mbox{input} & \mbox{output} \\ \hline
    \mbox{sync.} & $\mathbf{Q}$ & $\mathbf{P}$  \\ \hline
    \mbox{async.} & $\mathbf{M}$ & $\mathbf{N}$  \\
    \hline
  \end{tabular}$
\end{center}

The grammar of these types polarised with respect to input and output
occurrences is as follows; $Q$ and $P$ denote synchronous formulas in input and output position respectively, whereas $M$ and $N$ 
denote asynchronous formulas in input and output position respectively
(in the nonatomic case we will abbreviate thus: \emph{left sync., right synch., left async., and right async.}):
\disp{$
\begin{array}[t]{lrcl}
\mbox{Positive output} & P & ::= & 
\begin{array}[t]{l}
\mbox{At}^+\ |\ A\product B\out\ |\ I\out\ |\ A\swprod{k}B\out\ |\ J\out\ |\ A\adisj B\out
\end{array}\\
\mbox{Positive input} & Q & ::= & 
\begin{array}[t]{l}
\mbox{At}^-\ |\ C/B\inp\ |\ A\bsl C\inp\ |\ C\scircum{k}B\inp |\ A\sinfix{k}C\inp\ |\ A\aconj B\inp
\end{array}\\
\mbox{Negative output} & N & ::= &
\begin{array}[t]{l}
\mbox{At}^-\ |\ C/B\out\  |\ A\bsl C\out \ |\ C\scircum{k}B\out|\ A\sinfix{k}C\out\ |\ A\aconj B\out
\end{array} \\
\mbox{Negative input} & M & ::= & 
\begin{array}[t]{l}
\mbox{At}^+\ |\ A\product B\inp\ |\ I\inp\ |\ A\swprod{k}B\inp\ |\ J\inp\ |\ A\adisj B\inp
\end{array}\end{array}$}
Notice that if $P$ occurs in the antecedent then this occurrence of $P$ is negative,
and so forth. 

There are alternating phases of don't-care nondeterministic negative rule application,
and positive rule application locking on to \techterm{focalised\/} formulas.

Given a sequent with no occurrences of negative formulas, one chooses a positive formula
as principal formula (which is boxed;
we say it is focalised) and applies proof search to its subformulas while these remain positive. When one finds a negative formula or
a literal, invertible rules are applied in a don't care nondeterminitic fashion until
no longer possible, when another positive formula is chosen, and so on.

A sequent is either unfocused and as before, or else focused and has exactly one
type boxed.
The focalised logical rules are given in Figures~\ref{apmult}-\ref{rspadd}
including Curry-Howard categorial semantic labelling.
Occurrences of $P,Q,M$ and $N$ are supposed not to be focalised, which means that their
focalised occurrence \emph{must\/} be signalled with a box. By contrast, occurrences of $A,B,C$ may be focalised 
or not.

\begin{figure}
\begin{center}\mini
$\prooftree
\vect{A}\ass x, \Gamma\yields C\ass\chi
\justifies
\Gamma\yields A\bsl C\ass\lambda x\chi
\using \bsl R
\endprooftree$
\tb
$\prooftree
\Gamma, \vect{B}\ass y\yields C\ass\chi
\justifies
\Gamma\yields C/B\ass\lambda y\chi
\using / R
\endprooftree$
\vtab
$\prooftree
\Delta\langle\vect{A}\ass x, \vect{B}\ass y\rangle\yields D\ass\omega
\justifies
\Delta\langle\vect{A\product B}\ass z\rangle\yields D\ass\omega\{\pi_1 z/x, \pi_2 z/y\}
\using \product L
\endprooftree$ \vtab
$\prooftree
\Delta\langle\Lambda\rangle\yields A\ass\phi
\justifies
\Delta\langle\vect{I}\ass x\rangle\yields A\ass\phi
\using IL
\endprooftree$
\vtab
$\prooftree
\vect{A}\ass x\smwrap{k}\Gamma\yields C\ass\chi
\justifies
\Gamma\yields A\smwrap{k} C\ass\lambda x\chi
\using \sinfix{k} R
\endprooftree
\tb
\prooftree
\Gamma\smwrap{k}\vect{B}\ass y\yields C\ass\chi
\justifies
\Gamma\yields C\scircum{k} B\ass\lambda y\chi
\using \scircum{k} R
\endprooftree
$\vtab$
\prooftree
\Delta\langle\vect{A}\ass x\smwrap{k}\vect{B}\ass y\rangle\yields D\ass\omega
\justifies
\Delta\langle\vect{A\swprod{k} B}\ass z\rangle\yields D\ass\omega\subst{\pi_1 z/x, \pi_2 z/y}
\using \swprod{k} L
\endprooftree
$\vtab$
\prooftree
\Delta\langle\sep\rangle\yields A\ass\phi
\justifies
\Delta\langle\vect{J}\ass x\rangle\yields A\ass\phi
\using JL
\endprooftree
$
\end{center}
\caption{Asynchronous multiplicative rules}
\label{apmult}
\end{figure}

\begin{figure}
\begin{center}\mini
\prooftree
\Gamma\yields A\ass\phi\tb\Gamma\yields B\ass\psi
\justifies
\Gamma\yields A\aconj B\ass(\phi, \psi)
\using \aconj R
\endprooftree
\vtab
\prooftree
\Gamma\langle\vect{A}\ass x\rangle\yields C\ass\chi_1\tb\Gamma\langle\vect{B}\ass y\rangle\yields C\ass\chi_2
\justifies
\Gamma\langle\vect{A\adisj B}\ass z\rangle\yields C\ass z\rightarrow x.\chi_1; y.\chi_2
\using \adisj L
\endprooftree
\end{center}
\caption{Asynchronous additive rules}
\label{apadd}
\end{figure}


\begin{figure*}
\begin{center}\mini
$\prooftree
\Gamma\yields \mbox{\fbox{$P$}}\ass\phi \tb
\Delta\langle\vect{\mbox{\fbox{$Q$}}}\ass z\rangle\yields D\ass\omega
\justifies
\Delta\langle\Gamma, \vect{\mbox{\fbox{$P\bsl Q$}}}\ass y\rangle\yields D\ass\omega\{(y\ \phi)/z\}
\using \bsl L
\endprooftree \tb
\prooftree
\Gamma\yields \mbox{\fbox{$P$}}\ass\phi \tb
\Delta\langle\vect{M}\ass z\rangle\yields D\ass\omega
\justifies
\Delta\langle\Gamma, \vect{\mbox{\fbox{$P\bsl M$}}}\ass y\rangle\yields D\ass\omega\{(y\ \phi)/z\}
\using \bsl L
\endprooftree$
\vtab
$\prooftree
\Gamma\yields N\ass\phi \tb
\Delta\langle\vect{\mbox{\fbox{$Q$}}}\ass z\rangle\yields D\ass\omega
\justifies
\Delta\langle\Gamma, \vect{\mbox{\fbox{$N\bsl Q$}}}\ass y\rangle\yields D\ass\omega\{(y\ \phi)/z\}
\using \bsl L
\endprooftree \tb
\prooftree
\Gamma\yields N\ass\phi \tb
\Delta\langle\vect{M}\ass z\rangle\yields D\ass\omega
\justifies
\Delta\langle\Gamma, \vect{\mbox{\fbox{$N\bsl M$}}}\ass y\rangle\yields D\ass\omega\{(y\ \phi)/z\}
\using \bsl L
\endprooftree$
\vtab
$\prooftree
\Gamma\yields \mbox{\fbox{$P$}}\ass\psi \tb
\Delta\langle\vect{\mbox{\fbox{$Q$}}}\ass z\rangle\yields D\ass\omega
\justifies
\Delta\langle\vect{\mbox{\fbox{$Q/P$}}}\ass x, \Gamma\rangle\yields D\ass\omega\{(x\ \psi)/z\}
\using / L
\endprooftree \tb
\prooftree
\Gamma\yields N\ass\psi \tb
\Delta\langle\vect{\mbox{\fbox{$Q$}}}\ass z\rangle\yields D\ass\omega
\justifies
\Delta\langle\vect{\mbox{\fbox{$Q/N$}}}\ass x, \Gamma\rangle\yields D\ass\omega\{(x\ \psi)/z\}
\using / L
\endprooftree$
\vtab
$\prooftree
\Gamma\yields \mbox{\fbox{$P$}}\ass\psi \tb
\Delta\langle\vect{M}\ass z\rangle\yields D\ass\omega
\justifies
\Delta\langle\vect{\mbox{\fbox{$M/P$}}}\ass x, \Gamma\rangle\yields D\ass\omega\{(x\ \psi)/z\}
\using / L
\endprooftree \tb
\prooftree
\Gamma\yields M\ass\psi \tb
\Delta\langle\vect{N}\ass z\rangle\yields D\ass\omega
\justifies
\Delta\langle\vect{\mbox{\fbox{$N/M$}}}\ass x, \Gamma\rangle\yields D\ass\omega\{(x\ \psi)/z\}
\using / L
\endprooftree$
\commentout{\vtab
$
\prooftree
\Gamma_1\yields \mbox{\fbox{$P_1$}}\ass\phi\tb \Gamma_2\yields \mbox{\fbox{$P_2$}}\ass\psi
\justifies
\Gamma_1, \Gamma_2\yields \mbox{\fbox{$P_1\product P_2$}}\ass(\phi, \psi)
\using \product R
\endprooftree\tb
\prooftree
\Gamma_1\yields \mbox{\fbox{$P$}}\ass\phi\tb \Gamma_2\yields N\ass\psi
\justifies
\Gamma_1, \Gamma_2\yields \mbox{\fbox{$P\product N$}}\ass(\phi, \psi)
\using \product R
\endprooftree
$
\vtab
$
\prooftree
\Gamma_1\yields N\ass\phi\tb \Gamma_2\yields \mbox{\fbox{$P$}}\ass\psi
\justifies
\Gamma_1, \Gamma_2\yields \mbox{\fbox{$N\product P$}}\ass(\phi, \psi)
\using \product R
\endprooftree\tb
\prooftree
\Gamma_1\yields N_1\ass\phi\tb \Gamma_2\yields N_2\ass\psi
\justifies
\Gamma_1, \Gamma_2\yields \mbox{\fbox{$N_1\product N_2$}}\ass(\phi, \psi)
\using \product R
\endprooftree
$
\vtab
$
\prooftree
\justifies
\emptyset\sass\Lambda\yields \mbox{\fbox{$I$}}\ass t
\using IR
\endprooftree
$}
\end{center}
\caption{Left synchronous continuous multiplicative rules}
\label{spcmult}
\end{figure*}

\begin{figure*}
\begin{center}\mini
$\prooftree
\Gamma\yields \mbox{\fbox{$P$}}\ass\phi \tb
\Delta\langle\vect{\mbox{\fbox{$Q$}}}\ass z\rangle\yields D\ass\omega
\justifies
\Delta\langle\Gamma\smwrap{k}\vect{\mbox{\fbox{$P\sinfix{k} Q$}}}\ass y\rangle\yields D\ass\omega\{(y\ \phi)/z\}
\using \sinfix{k} L
\endprooftree \tb
\prooftree
\Gamma\yields \mbox{\fbox{$P$}}\ass\phi \tb
\Delta\langle\vect{M}\ass z\rangle\yields D\ass\omega
\justifies
\Delta\langle\Gamma\smwrap{k}\vect{\mbox{\fbox{$P\sinfix{k} M$}}}\ass y\rangle\yields D\ass\omega\{(y\ \phi)/z\}
\using \sinfix{k} L
\endprooftree$
\vtab
$\prooftree
\Gamma\yields N\ass\phi \tb
\Delta\langle\vect{\mbox{\fbox{$Q$}}}\ass z\rangle\yields D\ass\omega
\justifies
\Delta\langle\Gamma\smwrap{k}\vect{\mbox{\fbox{$N\sinfix{k} Q$}}}\ass y\rangle\yields D\ass\omega\{(y\ \phi)/z\}
\using \sinfix{k} L
\endprooftree \tb
\prooftree
\Gamma\yields N\ass\phi \tb
\Delta\langle\vect{M}\ass z\rangle\yields D\ass\omega
\justifies
\Delta\langle\Gamma\smwrap{k}\vect{\mbox{\fbox{$N\sinfix{k} M$}}}\ass y\rangle\yields D\ass\omega\{(y\ \phi)/z\}
\using \sinfix{k} L
\endprooftree$
\vtab
$\prooftree
\Gamma\yields \mbox{\fbox{$P$}}\ass\psi \tb
\Delta\langle\vect{\mbox{\fbox{$Q$}}}\ass z\rangle\yields D\ass\omega
\justifies
\Delta\langle\vect{\mbox{\fbox{$Q\scircum{k} P$}}}\ass x\smwrap{k}\Gamma\rangle\yields D\ass\omega\{(x\ \psi)/z\}
\using \scircum{k} L
\endprooftree \tb
\prooftree
\Gamma\yields N\ass\psi \tb
\Delta\langle\vect{\mbox{\fbox{$Q$}}}\ass z\rangle\yields D\ass\omega
\justifies
\Delta\langle\vect{\mbox{\fbox{$Q\scircum{k} N$}}}\ass x\smwrap{k}\Gamma\rangle\yields D\ass\omega\{(x\ \psi)/z\}
\using \scircum{k} L
\endprooftree
$
\vtab
$\prooftree
\Gamma\yields \mbox{\fbox{$P$}}\ass\psi \tb
\Delta\langle\vect{M}\ass z\rangle\yields D\ass\omega
\justifies
\Delta\langle\vect{\mbox{\fbox{$M\scircum{k} P$}}}\ass x\smwrap{k}\Gamma\rangle\yields D\ass\omega\{(x\ \psi)/z\}
\using \scircum{k} L
\endprooftree \tb
\prooftree
\Gamma\yields N\ass\psi \tb
\Delta\langle\vect{M}\ass z\rangle\yields D\ass\omega
\justifies
\Delta\langle\vect{\mbox{\fbox{$M\scircum{k} N$}}}\ass x\smwrap{k}\Gamma\rangle\yields D\ass\omega\{(x\ \psi)/z\}
\using \scircum{k} L
\endprooftree
$
\commentout{
\vtab
$
\prooftree
\Gamma_1\yields \mbox{\fbox{$P_1$}}\ass\phi\tb\Gamma_2\yields \mbox{\fbox{$P_2$}}\ass\psi
\justifies
\Gamma_1\smwrap{k}\Gamma_2\yields \mbox{\fbox{$P_1\swprod{k}P_2$}}\ass(\phi, \psi)
\using \swprod{k} R
\endprooftree\tb
\prooftree
\Omega\sass\Gamma_1\yields \mbox{\fbox{$P$}}\ass\phi\tb\Gamma_2\yields N\ass\psi
\justifies
\Gamma_1\smwrap{k}\Gamma_2\yields \mbox{\fbox{$P\swprod{k}N$}}\ass(\phi, \psi)
\using \swprod{k} R
\endprooftree$
\vtab
$
\prooftree
\Omega\sass\Gamma_1\yields N\ass\phi\tb\Gamma_2\yields \mbox{\fbox{$P$}}\ass\psi
\justifies
\Gamma_1\smwrap{k}\Gamma_2\yields \mbox{\fbox{$N\swprod{k}P$}}\ass(\phi, \psi)
\using \swprod{k} R
\endprooftree\tb
\prooftree
\Omega\sass\Gamma_1\yields N_1\ass\phi\tb\Gamma_2\yields N_2\ass\psi
\justifies
\Gamma_1\smwrap{k}\Gamma_2\yields \mbox{\fbox{$N_1\swprod{k}N_2$}}\ass(\phi, \psi)
\using \swprod{k} R
\endprooftree$
\vtab
$
\prooftree
\justifies
\emptyset\sass \sep\yields \mbox{\fbox{$J$}}\ass\zero
\using JR
\endprooftree
$}
\end{center}
\caption{Left synchronous discontinuous multiplicative rules}
\label{spimult}
\end{figure*}

\begin{figure}\mini
\begin{center}
\prooftree
\Gamma\langle\vect{\mbox{\fbox{$Q$}}}\ass x\rangle\yields C\ass\chi
\justifies
\Gamma\langle\vect{\mbox{\fbox{$Q\aconj B$}}}\ass z\rangle\yields C\ass\chi\subst{\pi_1 z/x}
\using \aconj L_1
\endprooftree
\tb
\prooftree
\Gamma\langle\vect{M}\ass x\rangle\yields C\ass\chi
\justifies
\Gamma\langle\vect{\mbox{\fbox{$M\aconj B$}}}\ass z\rangle\yields C\ass\chi\subst{\pi_1 z/x}
\using \aconj L_1
\endprooftree
\vtab
\prooftree
\Gamma\langle\vect{\mbox{\fbox{$Q$}}}\ass y\rangle\yields C\ass\chi
\justifies
\Gamma\langle\vect{\mbox{\fbox{$A\aconj Q$}}}\ass z\rangle\yields C\ass\chi\subst{\pi_2 z/y}
\using \aconj L_2
\endprooftree
\tb
\prooftree
\Gamma\langle\vect{M}\ass y\rangle\yields C\ass\chi
\justifies
\Gamma\langle\vect{\mbox{\fbox{$A\aconj M$}}}\ass z\rangle\yields C\ass\chi\subst{\pi_2 z/y}
\using \aconj L_2
\endprooftree
\end{center}
\caption{Left synchronous additive rules}
\label{spadd}
\end{figure}

\begin{figure*}
\begin{center}\mini
\prooftree
\Gamma_1\yields \mbox{\fbox{$P_1$}}\ass\phi\tb\Gamma_2\yields \mbox{\fbox{$P_2$}}\ass\psi
\justifies
\Gamma_1, \Gamma_2\yields \mbox{\fbox{$P_1\product P_2$}}\ass(\phi, \psi)
\using \product R
\endprooftree\tb
\prooftree
\Gamma_1\yields \mbox{\fbox{$P$}}\ass\phi\tb\Gamma_2\yields N\ass\psi
\justifies
\Gamma_1, \Gamma_2\yields \mbox{\fbox{$P\product N$}}\ass(\phi, \psi)
\using \product R
\endprooftree
\vtab
\prooftree
\Gamma_1\yields N\ass\phi\tb\Gamma_2\yields \mbox{\fbox{$P$}}\ass\psi
\justifies
\Gamma_1, \Gamma_2\yields \mbox{\fbox{$N\product P$}}\ass(\phi, \psi)
\using \product R
\endprooftree\tb
\prooftree
\Gamma_1\yields N_1\ass\phi\tb\Gamma_2\yields N_2\ass\psi
\justifies
\Gamma_1, \Gamma_2\yields \mbox{\fbox{$N_1\product N_2$}}\ass(\phi, \psi)
\using \product R
\endprooftree
\vtab
\prooftree
\justifies
\Lambda\yields \mbox{\fbox{$I$}}\ass\zero
\using IR
\endprooftree
\end{center}
\caption{Right synchronous continuous multiplicative rules}
\label{rspcmult}
\end{figure*}

\begin{figure*}
\begin{center}\mini
\prooftree
\Gamma_1\yields \mbox{\fbox{$P_1$}}\ass\phi\tb\Gamma_2\yields \mbox{\fbox{$P_2$}}\ass\psi
\justifies
\Gamma_1\smwrap{k}\Gamma_2\yields \mbox{\fbox{$P_1\swprod{k}P_2$}}\ass(\phi, \psi)
\using \swprod{k} R
\endprooftree\tb
\prooftree
\Gamma_1\yields \mbox{\fbox{$P$}}\ass\phi\tb\Gamma_2\yields N\ass\psi
\justifies
\Gamma_1\smwrap{k}\Gamma_2\yields \mbox{\fbox{$P\swprod{k}N$}}\ass(\phi, \psi)
\using \swprod{k} R
\endprooftree
\vtab
\prooftree
\Gamma_1\yields N\ass\phi\tb\Gamma_2\yields \mbox{\fbox{$P$}}\ass\psi
\justifies
\Gamma_1\smwrap{k}\Gamma_2\yields \mbox{\fbox{$N\swprod{k}P$}}\ass(\phi, \psi)
\using \swprod{k} R
\endprooftree\tb
\prooftree
\Gamma_1\yields N_1\ass\phi\tb\Gamma_2\yields N_2\ass\psi
\justifies
\Gamma_1\smwrap{k}\Gamma_2\yields \mbox{\fbox{$N_1\swprod{k}N_2$}}\ass(\phi, \psi)
\using \swprod{k} R
\endprooftree
\vtab
\prooftree
\justifies
\sep\yields \mbox{\fbox{$J$}}\ass\zero
\using JR
\endprooftree
\end{center}
\caption{Right synchronous discontinuous multiplicative rules}
\label{rspimult}
\end{figure*}

\begin{figure}\mini
\begin{center}
\prooftree
\Gamma\yields \mbox{\fbox{$P$}}\ass\phi
\justifies
\Gamma\yields \mbox{\fbox{$P\adisj B$}}\ass\iota_1\phi
\using \adisj R_1
\endprooftree
\tb
\prooftree
\Gamma\yields N\ass\phi
\justifies
\Gamma\yields \mbox{\fbox{$N\adisj B$}}\ass\iota_1\phi
\using \adisj R_1
\endprooftree
\vtab
\prooftree
\Gamma\yields \mbox{\fbox{$P$}}\ass\psi
\justifies
\Gamma \yields \mbox{\fbox{$A\adisj P$}}\ass\iota_2\psi
\using \adisj R_2
\endprooftree
\tb
\prooftree
\Gamma\yields N\ass\psi
\justifies
\Gamma \yields \mbox{\fbox{$A\adisj N$}}\ass\iota_2\psi
\using \adisj R_2
\endprooftree
\end{center}
\caption{Right synchronous additive rules}
\label{rspadd}
\end{figure}

\section{Completeness of focalisation for DA}
\label{compl}

We shall be dealing with three systems: the displacement calculus \DA{} with sequents notated $\Delta\yields A$, the \techterm{weakly focalised\/} displacement
calculus with additives \DAf{} with sequents notated $\Delta\lyieldsw A$, and the \techterm{strongly focalised\/} displacement calculus with additives \DAF{} with sequents notated
$\Delta\lyields A$. Sequents of both \DAf{} and \DAF{} may contain at most one focalised formula,
possibly $A$.
When a \DAf{} sequent is notated $\Delta\lyieldsw A\pf$, it means that the sequent possibly contains a (unique)  focalised formula. Otherwise, $\Delta\lyieldsw A$ means
that the sequent does not contain a focus. 

In this section we prove the strong focalisation property for the displacement
calculus with additives \DA{}.

The focalisation property for Linear Logic was discovered 
by \cite{andreoli:92}. In this paper we follow the proof idea from \cite{Laurent04bnote}, which we
adapt to the intuitionistic non-commutative case \DA{} with twin multiplicative modes of combination, the continuous (concatenation) and the discontinuous
(intercalation) products. The proof relies heavily on the Cut-elimination property for
weakly focalised \DA{} which is proved
in the appendix. In our presentation of focalisation
we have avoided the \techterm{react} rules of \cite{andreoli:92} and \cite{Chaudhuri:2006:FIM:1292706}, and use instead a simpler, box, notation 
suitable for non-commutativity.

\DAF{} is a subsystem of \DAf{}. \DAf{}
has the focusing rules \foc{} and Cut rules $\pcut_1$, $\pcut_2$, $\ncut_1$ and $\ncut_2$\footnote{If it is convenient,  we may drop 
the subscripts.} shown in (\ref{complfocD:defs}), and the synchronous and asynchronous rules displayed before, which are
read as allowing in synchronous rules the occurrence of asynchronous formulas, and in asynchronous rules as allowing arbitrary sequents with possibly
one focalised formula. 
\DAF{} has the focusing rules but not the Cut rules, and the synchronous and asynchronous rules displayed before, which are such that focalised
sequents cannot contain any complex asynchronous formulas, whereas sequents with at least one complex asynchronous formula cannot contain a
focalised formula. Hence, strongly focalised proof search operates in alternating asynchronous and synchronous phases. 
The weakly focalised calculus \DAf{} is an intermediate logic
which we use to prove the completeness of \DAF{} for \DA{}.
\disp{\mini$
\begin{array}[t]{l}
\prooftree
\Delta\langle \vect{\fbox{$Q$}}\rangle\lyieldsw A
\justifies
\Delta\langle \vect{Q}\rangle\lyieldsw A
\using \foc
\endprooftree
\tb
\prooftree
\Delta\lyieldsw \fbox{$P$}
\justifies
\Delta\lyieldsw P
\using \foc
\endprooftree\\\\
\prooftree
\Gamma\lyieldsw \fbox{$P$}
\tb 
\Delta\langle\vect P\rangle\lyieldsw C\pf 
\justifies
\Delta\langle\Gamma\rangle\lyieldsw C\pf
\using\pcut_1
\endprooftree
\tb
\prooftree
\Gamma\lyieldsw N\pf 
\tb 
\Delta\langle\vect{\fbox{$N$}}\rangle\lyieldsw C
\justifies
\Delta\langle\Gamma\rangle\lyieldsw C\pf
\using\pcut_2
\endprooftree\\\\
\prooftree
\Gamma\lyieldsw P\pf
\tb 
\Delta\langle\vect P\rangle\lyieldsw C
\justifies
\Delta\langle\Gamma\rangle\lyieldsw C\pf
\using\ncut_1
\endprooftree
\tb
\prooftree
\Gamma\lyieldsw N\
\tb \Delta\langle\vect{N}\rangle\lyieldsw C\pf
\justifies
\Delta\langle\Gamma\rangle\lyieldsw C\pf
\using \ncut_2
\endprooftree
\end{array}$}
\label{complfocD:defs}}


\subsection{Embedding of \DA{} into DA$_{\bf foc}$}

\label{complfocD:emb}

The identity axiom we consider for \DA{} and for both \DAf{} and \DAF{} is restricted to atomic types;
recalling that atomic types are classified into positive bias {\it At}$^+$ and negative bias {\it At}$^-$:
\disp{\begin{tabular}[t]{l}
If $P\in\mbox{\it At}^+$, P\lyieldsw \fbox{$P$} and P\lyields \fbox{$P$}\\
If $Q\in\mbox{\it At}^-$, \fbox{$Q$}\lyieldsw Q and \fbox{$Q$}\lyields Q\\
\end{tabular}}
 In fact, the Identity rule holds of any type A. 
It has the following formulation in the sequent calculi considered here:

\disp{\mini
$
\left\{
\begin{array}{lll}
\vect{A}\yields A&&\mbox{ in }\DA{}\\
\vect{P}\lyieldsw \fbox{$P$}&\tb \vect{\fbox{$N$}}\lyieldsw N&\mbox{ in }\DAf{}\\
\vect{P}\lyields P&\tb \vect{N}\lyields  N&\mbox{ in }\DAF{}\\
\end{array}\right.
$
}

The Identity axiom for arbitrary types is also known as \techterm{Eta-expansion}. Eta-expansion is easy to prove in both \DA{} and \DAf{}, 
but the same is not the case for \DAF{}. This is the reason to consider what we have called weak focalisation, which helps us to 
prove smoothly this crucial property for the proof of strong focalisation.

\begin{theorem}
\emph{(Embedding of \DA{} into \DAf{})}
For any configuration $\Delta$ and type $A$, we have that if $\Delta\yields A$ then $\Delta\lyieldsw A$.
\label{complfocD:embthm1}
\end{theorem}

\prf{
We proceed by induction on the length of the derivation of \DA{} proofs. In the following lines, we apply the induction hypothesis (i.h.) for each premise of \DA{} rules (with 
the exception of the Identity rule and the right rules of units):\\\\
- Identity axiom: 
\disp{\mini$
\prooftree
\vect{P}\lyieldsw \fbox{P}
\justifies
\vect{P}\lyieldsw P
\using \foc
\endprooftree
\tb
\prooftree
\vect{\fbox{N}}\lyieldsw N
\justifies
\vect{N}\lyieldsw N
\using \foc
\endprooftree
$}
- Cut rule: just apply \ncut.

\noindent
- Units
\disp{\mini$
\prooftree
\justifies
\Lambda\yields I
\using I R
\endprooftree
\tb\leadsto\tb
\prooftree
\prooftree
\justifies
\Lambda\lyieldsw \fbox{$I$}
\using IR
\endprooftree
\justifies
\Lambda\lyieldsw I
\using \foc
\endprooftree
$}
\disp{\mini$
\prooftree
\justifies
\sep\yields J
\using J R
\endprooftree
\tb\leadsto\tb
\prooftree
\prooftree
\justifies
\sep\lyieldsw \fbox{$J$}
\using JR
\endprooftree
\justifies
\sep\lyieldsw J
\using \foc
\endprooftree
$}
Left unit rules apply as in the case of \DA{}.\\\\

\noindent
- Left discontinuous product: directly translates.
%

\noindent
- Right discontinuous product. There are cases $P_1\swprod{k}P_2$, $N_1\swprod{k}N_2$, $N\swprod{k}P$ and $P\swprod{k}N$.
 We show one representative example:\\

\noindent{\mini$
\prooftree
\Delta\yields P
\tb 
\Gamma\yields N
\justifies
\Delta\smwrap{k}\Gamma\yields P\swprod{k}N
\using \swprod{k}R
\endprooftree
\tb\leadsto\tb$
$
\prooftree
\prooftree
\Gamma\lyieldsw N
\tb
\prooftree
\Delta\lyieldsw P
\tb
\prooftree
\vect{P}\lyieldsw \fbox{$P$}
\tb
\prooftree
\fbox{$\vect{N}$}\lyieldsw N
\justifies
\vect{N}\lyieldsw N
\using foc
\endprooftree
\justifies 
\vect{P}\smwrap{k}\vect{N}\yields \fbox{$P\swprod{k}N$}
\using\swprod{k}R
\endprooftree
\justifies
\Delta\smwrap{k}\vect{N}\lyieldsw \fbox{$P\swprod{k}N$}
\using \ncut_2
\endprooftree
\justifies
\Delta\smwrap{k}\Gamma\lyieldsw \fbox{$P\swprod{k}N$}
\using\ncut_2
\endprooftree
\justifies
\Delta\smwrap{k}\Gamma\lyieldsw P\swprod{k}N
\using foc
\endprooftree
$}\\

\noindent
- Left discontinuous \scircum{k} rule (the left rule for \sinfix{k} is entirely similar). Like in the case for the right discontinuous
product \swprod{k} rule, we only show one representative example:\\

\noindent{\mini$
\prooftree
\Gamma\yields P
\tb
\Delta\langle\vect{N}\rangle\yields A
\justifies 
\Delta\langle\vect{N\scircum{k}P}\smwrap{k}\Gamma\rangle\yields A
\using\scircum{k}L
\endprooftree
\tb\leadsto\tb$
$
\prooftree
\prooftree
\Gamma\lyieldsw P
\prooftree
\endprooftree
\tb
\prooftree
\prooftree
\vect{P}\lyieldsw\fbox{P}
\tb
\vect{\fbox{$N$}}\lyieldsw N
\justifies
\vect{\fbox{$N\scircum{k}P$}}\smwrap{k}\vect{P}\lyieldsw N
\using\scircum{k}L
\endprooftree
\tb
\Delta\langle\vect{N}\rangle\lyieldsw A
\justifies
\Delta\langle\vect{\fbox{$N\scircum{k}P$}}\smwrap{k}\vect{P}\rangle\lyieldsw A
\using \ncut_1
\endprooftree
\justifies
\Delta\langle\vect{\fbox{$N\scircum{k}P$}}\smwrap{k}\Gamma\rangle\lyieldsw A
\using \ncut_2
\endprooftree
\justifies 
\Delta\langle\vect{N\scircum{k}P}\smwrap{k}\Gamma\rangle\lyieldsw A
\using\foc
\endprooftree
$}\\

\noindent
- Right discontinuous \scircum{k} rule (the right discontinuous rule for \sinfix{k}{} is entirely similar):
\disp{\mini$
\prooftree
\Delta\smwrap{k}\vect{A}\yields B
\justifies
\Delta\yields B\scircum{k} A
\using \scircum{k} R
\endprooftree
\ \ \ \ \leadsto\ \ \ \ 
\prooftree
\Delta\smwrap{k}\vect{A}\lyieldsw B
\justifies
\Delta\lyieldsw B\scircum{k} A
\using \scircum{k} R
\endprooftree
$}
- Product and implicative continuous rules. These follow the same pattern as the discontinuous case.
We interchange the metalinguistic $k$-th intercalation $\smwrap{k}$  with the metalinguistic concatenation ',', and we
interchange \swprod{k}{}, \scircum{k}{} and \sinfix{k}{} with \product{}, $/$, and $\bsl$ respectively.

Concerning additives, conjunction Right translates directly and
we consider then conjunction Left (disjunction is symmetric):\\

\disp{\mini
\prooftree
\Delta\langle \vect{P}\rangle\yields C
\justifies
\Delta\langle \vect{P\aconj M}\rangle\yields C
\using\aconj L
\endprooftree
$\tb \leadsto\tb $
\prooftree
\vect{P\aconj M}\lyieldsw P\tb
\Delta\langle \vect{P}\rangle\lyieldsw C
\justifies
\Delta\langle \vect{P\aconj M}\rangle\lyieldsw C
\using\ncut_1
\endprooftree}

\noindent
where by Eta expansion and application of the \foc{} rule, we have $\vect{P\aconj M}\lyieldsw P$.
}

\subsection{Embedding of \DAf{} into \DAF{}}

\label{complfocD:emb2}

\begin{theorem}
\emph{(Embedding of \DAf{} into \DAF{})}
For any configuration $\Delta$ and type $A$, we have that if $\Delta\lyieldsw A$ with one focalised formula
and no asynchronous formula occurrence, then $\Delta\lyields A$ with the same formula focalised.
If $\Delta\lyieldsw A$ with no focalised formula and with at least one asynchronous formula, then 
$\Delta\lyields A$.
\label{complfocD:embthm2}
\end{theorem}

\prf{
We proceed by induction on the size of \DAf{} sequents.\footnote{For a given type $A$,
the \emph{size\/} of $A$, $|A|$, is the number of connectives in $A$. By recursion on configurations we have:
$
\begin{array}{rcl}
|\Lambda| & ::= & 0 \\
|\vect{A},\Delta| & ::= & |A| + |\Delta|, \mbox{for\ } sA=0\\
|\sep,\Delta| & ::= & |\Delta|\\
|A\{\Delta_1:\cdots:\Delta_{sA}\}|& ::= &|A|+\sum\limits_{i=1}^{sA}|\Delta_i|
\end{array}$

\noindent
Moreover, we have:

\noindent
$
|\Delta\langle\vect{\fbox{$Q$}}\rangle\lyieldsw A|=|\Delta\langle\vect{Q}\rangle\lyieldsw A|\\
|\Delta\lyieldsw \fbox{$P$}|=|\Delta\lyieldsw P|
$
} 
We consider Cut-free \DAf{} proofs which match the sequents of this theorem. If the last rule is logical (i.e., it is not an instance of the \foc{} rule) the 
i.h.\ applies
directly  and we get \DAF{} proofs of the same end-sequent. Now, let us suppose that the last rule is not logical, i.e.\ it is an instance of the \foc{} rule.
Let us suppose that the end sequent $\Delta\lyieldsw A$ is a synchronous sequent. Suppose for example that the focalised formula
is in the succedent:
\disp{\mini
$
\prooftree
\Delta\lyieldsw \fbox{$P$}
\justifies
\Delta\lyieldsw P
\using \foc
\endprooftree
$
}
The sequent $\Delta\lyieldsw \fbox{$P$}$ arises from a synchronous rule to which we can apply i.h..
Let us suppose now that the end-sequent contains at least one asynchronous formula. We see three cases which are illustrative:
\disp{\mini
\begin{tabular}[t]{ll}
a. & $\Delta\langle \vect{A\swprod{k}B}\rangle\lyieldsw \fbox{$P$}$\\
b. & $\Delta\langle \vect{\fbox{$Q$}}\rangle\lyieldsw B\scircum{k}A$\\
c. & $\Delta\langle \vect{\fbox{$Q$}}\rangle\lyieldsw A\aconj B$
\end{tabular}\label{complfocD:embthm2cases}}
We have by Eta expansion that $\vect{A\swprod{k}B}\lyieldsw \vect{\fbox{$A\swprod{k}B$}}$. We apply to this sequent the invertible
\swprod{k}{} left rule, whence $\vect{A}|_k\vect{B}\lyieldsw \vect{\fbox{$A\swprod{k}B$}}$. In case (\ref{complfocD:embthm2cases}a), 
we have the following proof in \DAf{}:
\disp{\mini
$
\prooftree
\prooftree
\vect{A}\smwrap{k}\vect{B}\lyieldsw \vect{\fbox{$A\swprod{k}B$}}
\tb
\Delta\langle \vect{A\swprod{k}B}\rangle\lyieldsw \fbox{$P$}
\justifies
\Delta\langle \vect{A}\smwrap{k}\vect{B}\rangle\lyieldsw \fbox{$P$}
\using\pcut_1
\endprooftree
\justifies
\Delta\langle \vect{A}\smwrap{k}\vect{B}\rangle\lyieldsw P
\using\foc
\endprooftree
$
}
To the above \DAf{} proof we apply Cut-elimination and we get the Cut-free \DAf{} end-sequent $\Delta\langle \vect{A}|_k\vect{B}\rangle$ $\lyieldsw P$. We have
$|\Delta\langle \vect{A}|_k\vect{B}\rangle\lyieldsw P| <  | \Delta\langle \vect{A\swprod{k}B}\rangle\lyieldsw P|$. We can apply then i.h.\ and we derive the provable
\DAF{} sequent $\Delta\langle \vect{A}\smwrap{k}\vect{B}\rangle\lyields P$ to which we can apply the left \swprod{k}{} rule. We have obtained 
$\Delta\langle \vect{A\swprod{k}B}\rangle\lyields P$. In the same way, we have that 
$\vect{\fbox{$B\scircum{k} A$}}|_k\vect{A}\lyieldsw B$.
Thus, in case (\ref{complfocD:embthm2cases}b), we have the following proof in \DAf{}:
\disp{\mini
$
\prooftree
\prooftree
\Delta\langle \vect{\fbox{$Q$}}\rangle\lyieldsw B\scircum{k}A
\tb
\vect{\fbox{$B\scircum{k} A$}}\smwrap{k}\vect{A}\lyieldsw B
\justifies
\Delta\langle \vect{\fbox{$Q$}}\rangle\smwrap{k}\vect{A}\lyieldsw B
\using\pcut_2
\endprooftree
\justifies
\Delta\langle \vect{Q}\rangle\smwrap{k}\vect{A}\lyieldsw B
\using\foc
\endprooftree
$
}
As before, we apply Cut-elimination to the above proof. We get the Cut-free \DAf{} end-sequent $\Delta\langle \vect{Q}\rangle|_k\vect{A}$ $\lyieldsw B$. It has
size less than $|\Delta\langle \vect{Q}\rangle\lyieldsw B\scircum{k}A|$. We can apply i.h.\ and we get the \DAF{} provable sequent 
$\Delta\langle \vect{Q}\rangle|_k\vect{A}\lyields B$ to which we apply the \scircum{k}{} right rule. 

In case (\ref{complfocD:embthm2cases}c):

\disp{
\prooftree
\Delta\langle \vect{\fbox{Q}}\rangle \lyieldsw A\aconj B
\justifies
\Delta\langle \vect{Q}\rangle \lyieldsw A\aconj B
\using\foc
\endprooftree}

\noindent
by applying the \foc{} rule and the invertibility of $\aconj R$
we get the provable \DAf{} sequents $\Delta\langle \vect{Q}\rangle \lyieldsw A$
and $\Delta\langle \vect{Q}\rangle \lyieldsw B$. These sequents have smaller size than $\Delta\langle \vect{Q}\rangle \lyieldsw A\aconj B$. The aforementioned sequents have a Cut-free proof in \DAf{}. We apply i.h.\ and we get $\Delta\langle \vect{Q}\rangle \lyields A$ and  $\Delta\langle \vect{Q}\rangle \lyields B$.
We apply the \aconj{} right rule in \DAF{}, and we get $\Delta\langle \vect{Q}\rangle \lyields A\aconj B$.
}\\

By this theorem we obtain the completeness of strong focalisation.

\section{Example}

We can have coordinate unlike types with nominal and
adjectival complementation of \emph{is}:
\disp{\mini
$[{\bf Tully}]{+}{\bf is}{+}[[{\bf Cicero}{+}{\bf and}{+}{\bf humanist}]]: Sf$}
Lexical lookup of types yields:
\disp{\mini
$[{\blacksquare}Nt(s(m)):{\it b}], {\blacksquare}(({\langle\rangle}{\exists}gNt(s(g))\backslash Sf)/({\exists}aNa{\oplus}({\exists}g({\it CN}{\it g}/{\it CN})): \\\lambda A\lambda B({\it Pres}\ ({\it A}\rightarrow C.[{\it B}={\it C}]; D.(({\it D}\ \lambda E[{\it E}={\it B}])\ {\it B}))),[[{\blacksquare}{\forall}gNt(s(g)): {\it 007},\\
 {\blacksquare}{\forall}f{\forall}a(({\blacksquare}((({\langle\rangle}Na\backslash Sf)/({\exists}bNb{\oplus}{\exists}g({\it CN}{\it g}/{\it CN}{\it g})))\backslash ({\langle\rangle}Na\backslash Sf))\backslash
  {[]^{-1}}{[]^{-1}}((({\langle\rangle}Na\backslash Sf)/\\({\exists}bNb{\oplus}{\exists}g({\it CN}{\it g}/{\it CN}))\backslash ({\langle\rangle}Na\backslash Sf)))/{\blacksquare}((({\langle\rangle}Na\backslash Sf)/({\exists}bNb{\oplus}{\exists}g({\it CN}{\it g}/{\it CN}{\it g})))\backslash ({\langle\rangle}Na\backslash Sf))):\\\lambda F\lambda G\lambda H\lambda I[(({\it G}\ {\it H})\ {\it I})\wedge (({\it F}\ {\it H})\ {\it I})], {\square}{\forall}n({\it CN}{\it n}/{\it CN}{\it n}): \mbox{\^{}}\lambda J\lambda K[({\it J}\ {\it K})\wedge (\mbox{\v{}}{\it teetotal}\ {\it K})]]]\ \Rightarrow\ Sf$}
The bracket modalities \mybrack{} and \abrack mark as syntactic domains subjects
and coordinate structures which are weak and strong islands respectively.
The quantifiers and first-order structure mark agreement features such as third person singular
for any gender for \syncnst{is}.
The normal modality $\square$ marks semantic intensionality and $\blacksquare$
marks rigid designator semantic intensionality.
The example has positive and negative additive disjunction
so that the derivation in Figures~\ref{cout1}--\ref{cout5} illustrates
both synchronous and asynchronous focusing additives.
This delivers the correct semantics:
$[({\it Pres}\ [{\it t}={\it c}])\wedge ({\it Pres}\ (\mbox{\v{}}{\it humanist}\ {\it t}))]$.
\begin{figure}
\begin{center}{
\prooftree
\prooftree
\prooftree
\prooftree
\prooftree
\prooftree
\prooftree
\prooftree
\prooftree
\prooftree
\prooftree
\prooftree
\justifies
{\it CN}{\it A}\ \Rightarrow\ {\it CN}{\it A}
\endprooftree
\prooftree
\justifies
\mbox{\fbox{${\it CN}{\it A}$}}\ \Rightarrow\ {\it CN}{\it A}
\endprooftree
\justifies
\mbox{\fbox{${\it CN}{\it A}/{\it CN}{\it A}$}}, {\it CN}{\it A}\ \Rightarrow\ {\it CN}{\it A}
\using {/}L
\endprooftree
\justifies
\mbox{\fbox{${\forall}n({\it CN}{\it n}/{\it CN}{\it n})$}}, {\it CN}{\it A}\ \Rightarrow\ {\it CN}{\it A}
\using {\forall}L
\endprooftree
\justifies
\mbox{\fbox{${\square}{\forall}n({\it CN}{\it n}/{\it CN}{\it n})$}}, {\it CN}{\it A}\ \Rightarrow\ {\it CN}{\it A}
\using {\Box}L
\endprooftree
\justifies
{\square}{\forall}n({\it CN}{\it n}/{\it CN}{\it n})\ \Rightarrow\ \fbox{${\it CN}{\it A}/{\it CN}{\it A}$}
\using {\sqcup}R
\endprooftree
\justifies
{\square}{\forall}n({\it CN}{\it n}/{\it CN}{\it n})\ \Rightarrow\ \fbox{${\exists}g({\it CN}{\it g}/{\it CN}{\it g})$}
\using {\exists}R
\endprooftree
\justifies
{\square}{\forall}n({\it CN}{\it n}/{\it CN}{\it n})\ \Rightarrow\ \fbox{${\exists}bNb{\oplus}{\exists}g({\it CN}{\it g}/{\it CN}{\it g})$}
\using {\oplus}R
\endprooftree
\prooftree
\prooftree
\prooftree
\justifies
Nt(s(m))\ \Rightarrow\ Nt(s(m))
\endprooftree
\justifies
[Nt(s(m))]\ \Rightarrow\ \fbox{${\langle\rangle}Nt(s(m))$}
\using {\langle\rangle}R
\endprooftree
\prooftree
\justifies
\mbox{\fbox{$Sf$}}\ \Rightarrow\ Sf
\endprooftree
\justifies
[Nt(s(m))], \mbox{\fbox{${\langle\rangle}Nt(s(m))\backslash Sf$}}\ \Rightarrow\ Sf
\using {\backslash}L
\endprooftree
\justifies
[Nt(s(m))], \mbox{\fbox{$({\langle\rangle}Nt(s(m))\backslash Sf)/({\exists}bNb{\oplus}{\exists}g({\it CN}{\it g}/{\it CN}{\it g}))$}}, {\square}{\forall}n({\it CN}{\it n}/{\it CN}{\it n})\ \Rightarrow\ Sf
\using {/}L
\endprooftree
\justifies
{\langle\rangle}Nt(s(m)), ({\langle\rangle}Nt(s(m))\backslash Sf)/({\exists}bNb{\oplus}{\exists}g({\it CN}{\it g}/{\it CN}{\it g})), {\square}{\forall}n({\it CN}{\it n}/{\it CN}{\it n})\ \Rightarrow\ Sf
\using {\langle\rangle}L
\endprooftree
\justifies
({\langle\rangle}Nt(s(m))\backslash Sf)/({\exists}bNb{\oplus}{\exists}g({\it CN}{\it g}/{\it CN}{\it g})), {\square}{\forall}n({\it CN}{\it n}/{\it CN}{\it n})\ \Rightarrow\ {\langle\rangle}Nt(s(m))\backslash Sf
\using {\backslash}R
\endprooftree
\justifies
{\square}{\forall}n({\it CN}{\it n}/{\it CN}{\it n})\ \Rightarrow\ (({\langle\rangle}Nt(s(m))\backslash Sf)/({\exists}bNb{\oplus}{\exists}g({\it CN}{\it g}/{\it CN}{\it g})))\backslash ({\langle\rangle}Nt(s(m))\backslash Sf)
\using {\backslash}R
\endprooftree
\justifies
\begin{array}{c}
{\square}{\forall}n({\it CN}{\it n}/{\it CN}{\it n})\ \Rightarrow\ {\blacksquare}((({\langle\rangle}Nt(s(m))\backslash Sf)/({\exists}bNb{\oplus}{\exists}g({\it CN}{\it g}/{\it CN}{\it g})))\backslash ({\langle\rangle}Nt(s(m))\backslash Sf))\\
\mbox{\footnotesize\textcircled{1}}
\end{array}
\using {\blacksquare}R
\endprooftree}
\end{center}
\caption{Coordination of unlike types, Part~I}
\label{cout1}
\end{figure}

\begin{figure}
\begin{center}{
\prooftree
\prooftree
\prooftree
\prooftree
\prooftree
\prooftree
\prooftree
\prooftree
\prooftree
\prooftree
\justifies
\mbox{\fbox{$Nt(s(A))$}}\ \Rightarrow\ Nt(s(A))
\endprooftree
\justifies
\mbox{\fbox{${\forall}gNt(s(g))$}}\ \Rightarrow\ Nt(s(A))
\using {\forall}L
\endprooftree
\justifies
\mbox{\fbox{${\blacksquare}{\forall}gNt(s(g))$}}\ \Rightarrow\ Nt(s(A))
\using {\blacksquare}L
\endprooftree
\justifies
{\blacksquare}{\forall}gNt(s(g))\ \Rightarrow\ \fbox{${\exists}bNb$}
\using {\exists}R
\endprooftree
\justifies
{\blacksquare}{\forall}gNt(s(g))\ \Rightarrow\ \fbox{${\exists}bNb{\oplus}{\exists}g({\it CN}{\it g}/{\it CN}{\it g})$}
\using {\oplus}R
\endprooftree
\prooftree
\prooftree
\prooftree
\justifies
Nt(s(m))\ \Rightarrow\ Nt(s(m))
\endprooftree
\justifies
[Nt(s(m))]\ \Rightarrow\ \fbox{${\langle\rangle}Nt(s(m))$}
\using {\langle\rangle}R
\endprooftree
\prooftree
\justifies
\mbox{\fbox{$Sf$}}\ \Rightarrow\ Sf
\endprooftree
\justifies
[Nt(s(m))], \mbox{\fbox{${\langle\rangle}Nt(s(m))\backslash Sf$}}\ \Rightarrow\ Sf
\using {\backslash}L
\endprooftree
\justifies
[Nt(s(m))], \mbox{\fbox{$({\langle\rangle}Nt(s(m))\backslash Sf)/({\exists}bNb{\oplus}{\exists}g({\it CN}{\it g}/{\it CN}{\it g}))$}}, {\blacksquare}{\forall}gNt(s(g))\ \Rightarrow\ Sf
\using {/}L
\endprooftree
\justifies
{\langle\rangle}Nt(s(m)), ({\langle\rangle}Nt(s(m))\backslash Sf)/({\exists}bNb{\oplus}{\exists}g({\it CN}{\it g}/{\it CN}{\it g})), {\blacksquare}{\forall}gNt(s(g))\ \Rightarrow\ Sf
\using {\langle\rangle}L
\endprooftree
\justifies
({\langle\rangle}Nt(s(m))\backslash Sf)/({\exists}bNb{\oplus}{\exists}g({\it CN}{\it g}/{\it CN}{\it g})), {\blacksquare}{\forall}gNt(s(g))\ \Rightarrow\ {\langle\rangle}Nt(s(m))\backslash Sf
\using {\backslash}R
\endprooftree
\justifies
{\blacksquare}{\forall}gNt(s(g))\ \Rightarrow\ (({\langle\rangle}Nt(s(m))\backslash Sf)/({\exists}bNb{\oplus}{\exists}g({\it CN}{\it g}/{\it CN}{\it g})))\backslash ({\langle\rangle}Nt(s(m))\backslash Sf)
\using {\backslash}R
\endprooftree
\justifies
\begin{array}{c}
{\blacksquare}{\forall}gNt(s(g))\ \Rightarrow\ {\blacksquare}((({\langle\rangle}Nt(s(m))\backslash Sf)/({\exists}bNb{\oplus}{\exists}g({\it CN}{\it g}/{\it CN}{\it g})))\backslash ({\langle\rangle}Nt(s(m))\backslash Sf))\\
\mbox{\footnotesize\textcircled{2}}
\end{array}
\using {\blacksquare}R
\endprooftree}
\end{center}
\caption{Coordination of unlike types, Part~II}
\label{cout2}
\end{figure}

\begin{figure}
\begin{center}{
\prooftree
\prooftree
\prooftree
\prooftree
\prooftree
\prooftree
\prooftree
\justifies
N2\ \Rightarrow\ N2
\endprooftree
\justifies
N2\ \Rightarrow\ \fbox{${\exists}aNa$}
\using {\exists}R
\endprooftree
\justifies
N2\ \Rightarrow\ \fbox{${\exists}aNa{\oplus}{\exists}g({\it CN}{\it g}/{\it CN}{\it g})$}
\using {\oplus}R
\endprooftree
\prooftree
\prooftree
\prooftree
\prooftree
\justifies
Nt(s(m))\ \Rightarrow\ Nt(s(m))
\endprooftree
\justifies
Nt(s(m))\ \Rightarrow\ \fbox{${\exists}gNt(s(g))$}
\using {\exists}R
\endprooftree
\justifies
[Nt(s(m))]\ \Rightarrow\ \fbox{${\langle\rangle}{\exists}gNt(s(g))$}
\using {\langle\rangle}R
\endprooftree
\prooftree
\justifies
\mbox{\fbox{$Sf$}}\ \Rightarrow\ Sf
\endprooftree
\justifies
[Nt(s(m))], \mbox{\fbox{${\langle\rangle}{\exists}gNt(s(g))\backslash Sf$}}\ \Rightarrow\ Sf
\using {\backslash}L
\endprooftree
\justifies
[Nt(s(m))], \mbox{\fbox{$({\langle\rangle}{\exists}gNt(s(g))\backslash Sf)/({\exists}aNa{\oplus}{\exists}g({\it CN}{\it g}/{\it CN}{\it g})$}}, N2\ \Rightarrow\ Sf
\using {/}L
\endprooftree
\justifies
[Nt(s(m))], \mbox{\fbox{${\blacksquare}(({\langle\rangle}{\exists}gNt(s(g))\backslash Sf)/({\exists}aNa{\oplus}{\exists}g({\it CN}{\it g}/{\it CN}{\it g})))$}}, N2\ \Rightarrow\ Sf
\using {\blacksquare}L
\endprooftree
\justifies
[Nt(s(m))], {\blacksquare}(({\langle\rangle}{\exists}gNt(s(g))\backslash Sf)/({\exists}aNa{\oplus}{\exists}g({\it CN}{\it g}/{\it CN}{\it g}))), {\exists}bNb\ \Rightarrow\ Sf
\using {\exists}L
\endprooftree
\justifies
\begin{array}{c}
{\langle\rangle}Nt(s(m)), {\blacksquare}(({\langle\rangle}{\exists}gNt(s(g))\backslash Sf)/({\exists}aNa{\oplus}{\exists}g({\it CN}{\it g}/{\it CN}{\it g}))), {\exists}bNb\ \Rightarrow\ Sf\\
\mbox{\footnotesize\textcircled{3}}
\end{array}
\using {\langle\rangle}L
\endprooftree
}
\end{center}
\caption{Coordination of unlike types, Part III}
\label{cout3}
\end{figure}

\begin{figure}
\begin{center}{
\prooftree
\prooftree
\prooftree
\prooftree
\prooftree
\prooftree
\prooftree
\prooftree
\prooftree
\prooftree
\prooftree
\justifies
{\it CN}{\it 1}\ \Rightarrow\ {\it CN}{\it 1}
\endprooftree
\prooftree
\justifies
\mbox{\fbox{${\it CN}{\it 1}$}}\ \Rightarrow\ {\it CN}{\it 1}
\endprooftree
\justifies
\mbox{\fbox{${\it CN}{\it 1}/{\it CN}{\it 1}$}}, {\it CN}{\it 1}\ \Rightarrow\ {\it CN}{\it 1}
\using {/}L
\endprooftree
\justifies
{\it CN}{\it 1}/{\it CN}{\it 1}\ \Rightarrow\ {\it CN}{\it 1}/{\it CN}{\it 1}
\using {/}R
\endprooftree
\justifies
{\it CN}{\it 1}/{\it CN}{\it 1}\ \Rightarrow\ \fbox{$({\it CN}{\it 1}/{\it CN}{\it 1}){\sqcup}({\it CN}{\it 1}\backslash {\it CN}{\it 1})$}
\using {\sqcup}R
\endprooftree
\justifies
{\it CN}{\it 1}/{\it CN}{\it 1}\ \Rightarrow\ \fbox{${\exists}g(({\it CN}{\it g}/{\it CN}{\it g}){\sqcup}({\it CN}{\it g}\backslash {\it CN}{\it g}))$}
\using {\exists}R
\endprooftree
\justifies
{\it CN}{\it 1}/{\it CN}{\it 1}\ \Rightarrow\ \fbox{${\exists}g(({\it CN}{\it g}/{\it CN}{\it g}){\sqcup}({\it CN}{\it g}\backslash {\it CN}{\it g}))$}
\using {-}R
\endprooftree
\justifies
{\it CN}{\it 1}/{\it CN}{\it 1}\ \Rightarrow\ \fbox{${\exists}aNa{\oplus}({\exists}g(({\it CN}{\it g}/{\it CN}{\it g}){\sqcup}({\it CN}{\it g}\backslash {\it CN}{\it g})){-}I)$}
\using {\oplus}R
\endprooftree
\prooftree
\prooftree
\prooftree
\prooftree
\justifies
Nt(s(m))\ \Rightarrow\ Nt(s(m))
\endprooftree
\justifies
Nt(s(m))\ \Rightarrow\ \fbox{${\exists}gNt(s(g))$}
\using {\exists}R
\endprooftree
\justifies
[Nt(s(m))]\ \Rightarrow\ \fbox{${\langle\rangle}{\exists}gNt(s(g))$}
\using {\langle\rangle}R
\endprooftree
\prooftree
\justifies
\mbox{\fbox{$Sf$}}\ \Rightarrow\ Sf
\endprooftree
\justifies
[Nt(s(m))], \mbox{\fbox{${\langle\rangle}{\exists}gNt(s(g))\backslash Sf$}}\ \Rightarrow\ Sf
\using {\backslash}L
\endprooftree
\justifies
[Nt(s(m))], \mbox{\fbox{$({\langle\rangle}{\exists}gNt(s(g))\backslash Sf)/({\exists}aNa{\oplus}({\exists}g(({\it CN}{\it g}/{\it CN}{\it g}){\sqcup}({\it CN}{\it g}\backslash {\it CN}{\it g})){-}I))$}}, {\it CN}{\it 1}/{\it CN}{\it 1}\ \Rightarrow\ Sf
\using {/}L
\endprooftree
\justifies
[Nt(s(m))], \mbox{\fbox{${\blacksquare}(({\langle\rangle}{\exists}gNt(s(g))\backslash Sf)/({\exists}aNa{\oplus}{\exists}g(({\it CN}{\it g}/{\it CN}{\it g})))$}}, {\it CN}{\it 1}/{\it CN}{\it 1}\ \Rightarrow\ Sf
\using {\blacksquare}L
\endprooftree
\justifies
[Nt(s(m))], {\blacksquare}(({\langle\rangle}{\exists}gNt(s(g))\backslash Sf)/({\exists}aNa{\oplus}{\exists}g({\it CN}{\it g}/{\it CN}{\it g}))), {\exists}g({\it CN}{\it g}/{\it CN}{\it g})\ \Rightarrow\ Sf
\using {\exists}L
\endprooftree
\justifies
\begin{array}{c}
{\langle\rangle}Nt(s(m)), {\blacksquare}(({\langle\rangle}{\exists}gNt(s(g))\backslash Sf)/({\exists}aNa{\oplus}{\exists}g({\it CN}{\it g}/{\it CN}{\it g}))), {\exists}g({\it CN}{\it g}/{\it CN}{\it g})\ \Rightarrow\ Sf\\
\mbox{\footnotesize\textcircled{4}}
\end{array}
\using {\langle\rangle}L
\endprooftree}
\end{center}
\caption{Coordination of unlike types, Part IV}
\label{cout4}
\end{figure}

\begin{figure}
\begin{center}
$\begin{array}{c}
\rotatebox{90}{\scriptsize
\prooftree
\prooftree
\prooftree
\prooftree
\mbox{\footnotesize\textcircled{1}}\tab
\prooftree
\mbox{\footnotesize\textcircled{2}}\tab
\prooftree
\prooftree
\prooftree
\prooftree
\prooftree
\prooftree
\mbox{\footnotesize\textcircled{3}}\tab
\mbox{\footnotesize\textcircled{4}}
\justifies
{\langle\rangle}Nt(s(m)), {\blacksquare}(({\langle\rangle}{\exists}gNt(s(g))\backslash Sf)/({\exists}aNa{\oplus}{\exists}g({\it CN}{\it g}/{\it CN}{\it g}))), {\exists}bNb{\oplus}{\exists}g({\it CN}{\it g}/{\it CN}{\it g})\ \Rightarrow\ Sf
\using {\oplus}L
\endprooftree
\justifies
{\blacksquare}(({\langle\rangle}{\exists}gNt(s(g))\backslash Sf)/({\exists}aNa{\oplus}{\exists}g({\it CN}{\it g}/{\it CN}{\it g}))), {\exists}bNb{\oplus}{\exists}g({\it CN}{\it g}/{\it CN}{\it g})\ \Rightarrow\ {\langle\rangle}Nt(s(m))\backslash Sf
\using {\backslash}R
\endprooftree
\justifies
{\blacksquare}(({\langle\rangle}{\exists}gNt(s(g))\backslash Sf)/({\exists}aNa{\oplus}{\exists}g({\it CN}{\it g}/{\it CN}{\it g}))\ \Rightarrow\ ({\langle\rangle}Nt(s(m))\backslash Sf)/({\exists}bNb{\oplus}{\exists}g({\it CN}{\it g}/{\it CN}{\it g}))
\using {/}R
\endprooftree
\prooftree
\prooftree
\prooftree
\prooftree
\justifies
\mbox{\fbox{$Nt(s(m))$}}\ \Rightarrow\ Nt(s(m))
\endprooftree
\justifies
\mbox{\fbox{${\blacksquare}Nt(s(m))$}}\ \Rightarrow\ Nt(s(m))
\using {\blacksquare}L
\endprooftree
\justifies
[{\blacksquare}Nt(s(m))]\ \Rightarrow\ \fbox{${\langle\rangle}Nt(s(m))$}
\using {\langle\rangle}R
\endprooftree
\prooftree
\justifies
\mbox{\fbox{$Sf$}}\ \Rightarrow\ Sf
\endprooftree
\justifies
[{\blacksquare}Nt(s(m))], \mbox{\fbox{${\langle\rangle}Nt(s(m))\backslash Sf$}}\ \Rightarrow\ Sf
\using {\backslash}L
\endprooftree
\justifies
[{\blacksquare}Nt(s(m))], {\blacksquare}(({\langle\rangle}{\exists}gNt(s(g))\backslash Sf)/({\exists}aNa{\oplus}{\exists}g({\it CN}{\it g}/{\it CN}{\it g}))), \mbox{\fbox{$(({\langle\rangle}Nt(s(m))\backslash Sf)/({\exists}bNb{\oplus}{\exists}g({\it CN}{\it g}/{\it CN}{\it g})))\backslash ({\langle\rangle}Nt(s(m))\backslash Sf)$}}\ \Rightarrow\ Sf
\using {\backslash}L
\endprooftree
\justifies
[{\blacksquare}Nt(s(m))], {\blacksquare}(({\langle\rangle}{\exists}gNt(s(g))\backslash Sf)/({\exists}aNa{\oplus}{\exists}g({\it CN}{\it g}/{\it CN}{\it g}))), [\mbox{\fbox{${[]^{-1}}((({\langle\rangle}Nt(s(m))\backslash Sf)/({\exists}bNb{\oplus}{\exists}g({\it CN}{\it g}/{\it CN}{\it g}))))\backslash ({\langle\rangle}Nt(s(m))\backslash Sf))$}}]\ \Rightarrow\ Sf
\using {[]^{-1}}L
\endprooftree
\justifies
[{\blacksquare}Nt(s(m))], {\blacksquare}(({\langle\rangle}{\exists}gNt(s(g))\backslash Sf)/({\exists}aNa{\oplus}{\exists}g({\it CN}{\it g}/{\it CN}{\it g}))), [[\mbox{\fbox{${[]^{-1}}{[]^{-1}}((({\langle\rangle}Nt(s(m))\backslash Sf)/({\exists}bNb{\oplus}{\exists}g({\it CN}{\it g}/{\it CN}{\it g}))))\backslash ({\langle\rangle}Nt(s(m))\backslash Sf))$}}]]\ \Rightarrow\ Sf
\using {[]^{-1}}L
\endprooftree
\justifies
\begin{array}{c}
{}[{\blacksquare}Nt(s(m))], {\blacksquare}(({\langle\rangle}{\exists}gNt(s(g))\backslash Sf)/({\exists}aNa{\oplus}{\exists}g({\it CN}{\it g}/{\it CN}{\it g}))), [[{\blacksquare}{\forall}gNt(s(g)),\\
\mbox{\fbox{${\blacksquare}((({\langle\rangle}Nt(s(m))\backslash Sf)/({\exists}bNb{\oplus}{\exists}g({\it CN}{\it g}/{\it CN}{\it g})))\backslash ({\langle\rangle}Nt(s(m))\backslash Sf))\backslash {[]^{-1}}{[]^{-1}}((({\langle\rangle}Nt(s(m))\backslash Sf)/({\exists}bNb{\oplus}{\exists}g({\it CN}{\it g}/{\it CN}{\it g})))\backslash ({\langle\rangle}Nt(s(m))\backslash Sf))$}}]]\ \Rightarrow\ Sf
\end{array}
\using {\backslash}L
\endprooftree
\justifies
\begin{array}{c}
{}[{\blacksquare}Nt(s(m))], {\blacksquare}(({\langle\rangle}{\exists}gNt(s(g))\backslash Sf)/({\exists}aNa{\oplus}{\exists}g({\it CN}{\it g}/{\it CN}{\it g}))), [[{\blacksquare}{\forall}gNt(s(g)),\\
\mbox{\fbox{$({\blacksquare}((({\langle\rangle}Nt(s(m))\backslash Sf)/({\exists}bNb{\oplus}{\exists}g({\it CN}{\it g}/{\it CN}{\it g})))\backslash ({\langle\rangle}Nt(s(m))\backslash Sf))\backslash {[]^{-1}}{[]^{-1}}((({\langle\rangle}Nt(s(m))\backslash Sf)/({\exists}bNb{\oplus}{\exists}g({\it CN}{\it g}/{\it CN}{\it g})))\backslash ({\langle\rangle}Nt(s(m))\backslash Sf)))/{\blacksquare}((({\langle\rangle}Nt(s(m))\backslash Sf)/({\exists}bNb{\oplus}{\exists}g({\it CN}{\it g}/{\it CN}{\it g})))\backslash ({\langle\rangle}Nt(s(m))\backslash Sf))$}},\\
{\square}{\forall}n({\it CN}{\it n}/{\it CN}{\it n})]]\ \Rightarrow\ Sf
\end{array}
\using {/}L
\endprooftree
\justifies
\begin{array}{c}
{}[{\blacksquare}Nt(s(m))], {\blacksquare}(({\langle\rangle}{\exists}gNt(s(g))\backslash Sf)/({\exists}aNa{\oplus}{\exists}g({\it CN}{\it g}/{\it CN}{\it g}))), [[{\blacksquare}{\forall}gNt(s(g)),\\
\mbox{\fbox{${\forall}a(({\blacksquare}((({\langle\rangle}Na\backslash Sf)/({\exists}bNb{\oplus}{\exists}g({\it CN}{\it g}/{\it CN}{\it g})))\backslash ({\langle\rangle}Na\backslash Sf))\backslash {[]^{-1}}{[]^{-1}}((({\langle\rangle}Na\backslash Sf)/({\exists}bNb{\oplus}{\exists}g({\it CN}{\it g}/{\it CN}{\it g})))\backslash ({\langle\rangle}Na\backslash Sf)))/{\blacksquare}((({\langle\rangle}Na\backslash Sf)/({\exists}bNb{\oplus}{\exists}g({\it CN}{\it g}/{\it CN}{\it g})))\backslash ({\langle\rangle}Na\backslash Sf)))$}},\\
{\square}{\forall}n({\it CN}{\it n}/{\it CN}{\it n})]]\ \Rightarrow\ Sf
\end{array}
\using {\forall}L
\endprooftree
\justifies
\begin{array}{c}
{}[{\blacksquare}Nt(s(m))], {\blacksquare}(({\langle\rangle}{\exists}gNt(s(g))\backslash Sf)/({\exists}aNa{\oplus}{\exists}g({\it CN}{\it g}/{\it CN}{\it g}))), [[{\blacksquare}{\forall}gNt(s(g)),\\
\mbox{\fbox{${\forall}f{\forall}a(({\blacksquare}((({\langle\rangle}Na\backslash Sf)/({\exists}bNb{\oplus}{\exists}g({\it CN}{\it g}/{\it CN}{\it g})))\backslash ({\langle\rangle}Na\backslash Sf))\backslash {[]^{-1}}{[]^{-1}}((({\langle\rangle}Na\backslash Sf)/({\exists}bNb{\oplus}{\exists}g({\it CN}{\it g}/{\it CN}{\it g}))))\backslash ({\langle\rangle}Na\backslash Sf)))/{\blacksquare}((({\langle\rangle}Na\backslash Sf)/({\exists}bNb{\oplus}{\exists}g({\it CN}{\it g}/{\it CN}{\it g})))\backslash ({\langle\rangle}Na\backslash Sf)))$}},\\
{\square}{\forall}n({\it CN}{\it n}/{\it CN}{\it n})]]\ \Rightarrow\ Sf
\end{array}
\using {\forall}L
\endprooftree
\justifies
\begin{array}{c}
{}[{\blacksquare}Nt(s(m))], {\blacksquare}(({\langle\rangle}{\exists}gNt(s(g))\backslash Sf)/({\exists}aNa{\oplus}{\exists}g({\it CN}{\it g}/{\it CN}{\it g}))), [[{\blacksquare}{\forall}gNt(s(g)),\\
\mbox{\fbox{${\blacksquare}{\forall}f{\forall}a(({\blacksquare}((({\langle\rangle}Na\backslash Sf)/({\exists}bNb{\oplus}{\exists}g({\it CN}{\it g}/{\it CN}{\it g})))\backslash ({\langle\rangle}Na\backslash Sf))\backslash {[]^{-1}}{[]^{-1}}((({\langle\rangle}Na\backslash Sf)/({\exists}bNb{\oplus}{\exists}g({\it CN}{\it g}/{\it CN}{\it g})))\backslash ({\langle\rangle}Na\backslash Sf)))/{\blacksquare}((({\langle\rangle}Na\backslash Sf)/({\exists}bNb{\oplus}{\exists}g({\it CN}{\it g}/{\it CN}{\it g})))\backslash ({\langle\rangle}Na\backslash Sf)))$}},\\
{\square}{\forall}n({\it CN}{\it n}/{\it CN}{\it n})]]\ \Rightarrow\ Sf
\end{array}
\using {\blacksquare}L
\endprooftree}
\end{array}$
\end{center}
\caption{Coordination of unlike types, Part~V}
\label{cout5}
\end{figure}

\clearpage

{

\bibliographystyle{eptcs}
\bibliography{bib151104}
}

\section*{Appendix: Cut Elimination}
We prove this by induction on the complexity $(d,h)$ of top-most instances of $Cut$, where $d$ is the size\footnote{The size of $|A|$ is the number of connectives appearing
in $A$.} of the cut formula and $h$ is the length of the derivation the last rule of which is the Cut rule.
There are four cases to consider: Cut with axiom in the minor premise, Cut with axiom in the major premise, principal Cuts, and
permutation conversions. In each case, the complexity of the Cut is reduced. In order to save space, we will not be exhaustive showing all the cases
because many follow the same pattern. In particular, for any synchronous logical rule there are always four cases to consider corresponding to 
the polarity of the subformulas. Here, and in the following, we will show only one representative example. 
Concerning continuous and discontinuous formulas, we will show only the discontinuous cases
(discontinuous connectives are less known than the continuous ones of the plain Lambek Calculus). For the continuous instances, the reader has only to interchange
the meta-linguistic wrap $|_k$ with the meta-linguistic concatenation $','$, \swprod{k}{} with \product{}, \scircum{k} with $/$ and 
\sinfix{k}{} with \bsl{}. The units cases (principal case and permutation conversion cases) are completely trivial.\\\\
\prf{
- $Id$ cases: 
\disp{
\prooftree
\vect{P}\lyieldsw\fbox{$P$}\tb 
\Delta\langle\vect{P}\rangle\lyieldsw B\pf
\justifies
\Delta\langle\vect{P}\rangle\lyieldsw B\pf
\using\pcut_1
\endprooftree
\tb$\leadsto$\tb
$\Delta\langle\vect{P}\rangle\lyieldsw B$\pf
\tb \\\\\\
\prooftree
\Delta\lyieldsw N\pf
\tb
\vect{\fbox{$N$}}\lyieldsw N
\justifies
\Delta\langle\vect{N}\rangle\lyieldsw B\pf
\using\pcut_2
\endprooftree
\tb$\leadsto$\tb
$\Delta\langle\vect{N}\rangle\lyieldsw B\pf$
}
The attentive reader may have wondered whether the following $Id$ case could arise:\\
\disp{
\prooftree
\fbox{$\vect{Q}$}\yields Q
\tb
\Gamma\langle\vect{Q}\rangle\yields A
\justifies
\Gamma\langle\fbox{Q}\rangle\yields A
\using \ncut_i
\endprooftree
\label{patholCut1}
}
If $Q$ were a primitive type $q$, and $\Gamma$ were not the empty context, we would have then a Cut-free 
underivable sequent. For example, if the right premise of the Cut rule in~(\ref{patholCut1})
were the derivable sequent $q,q\bsl s\yields s$, we would have then as conclusion:
\disp{
$
\fbox{q},q\bsl s\yields s
$\label{patholCutEx}}
Since the primitive type $q$ in the antecedent is focalised, there is no possibility of applying the
\bsl{} left rule, which is a synchronous rule that needs that its active formula to be focalised. 
Principal cases:\\\\
\product{} \foc{} cases:
\disp{ 
\prooftree
\prooftree
\Delta\lyieldsw \fbox{$P$}
\justifies
\Delta\lyieldsw P
\using\foc
\endprooftree
\tb
\Gamma\langle\vect{P}\rangle\lyieldsw A\pf
\justifies
\Gamma\langle\Delta\rangle\lyieldsw A\pf
\using\ncut_1
\endprooftree
\tb$\leadsto$\tb
\prooftree
\Delta\lyieldsw \fbox{$P$}
\tb
\Gamma\langle\vect{P}\rangle\lyieldsw A\pf
\justifies
\Gamma\langle\Delta\rangle\lyieldsw A\pf
\using\pcut_1
\endprooftree
}
\disp{ 
\prooftree
\Delta\lyieldsw N
\tb
\prooftree
\Delta\langle\fbox{$N$}\rangle\lyieldsw A
\justifies
\Gamma\langle\vect{N}\rangle\lyieldsw A
\using\foc
\endprooftree
\justifies
\Gamma\langle\Delta\rangle\lyieldsw A
\using\ncut_2
\endprooftree
\tb$\leadsto$\tb
\prooftree
\Delta\lyieldsw N
\tb
\Gamma\langle\vect{\fbox{$N$}}\rangle\lyieldsw A
\justifies
\Gamma\langle\Delta\rangle\lyieldsw A
\using\pcut_2
\endprooftree
}
\product{} logical connectives:
\disp{
\prooftree
\prooftree
\Delta|_k\vect{P_1}\lyieldsw P_2\pf
\justifies
\Delta\lyieldsw P_2\scircum{k}P_1\pf
\using\scircum{k}R
\endprooftree
\tb
\prooftree
\Gamma_1\lyieldsw \fbox{$P_1$}
\tb
\Gamma_2\langle\vect{P_2}\rangle\lyieldsw A
\justifies
\Gamma_2\langle\vect{\fbox{$P_2\scircum{k}P_1$}}|_k\Gamma_1\rangle\lyieldsw A
\using\scircum{k}L
\endprooftree
\justifies
\Gamma_2\langle\Delta|_k\Gamma_1\rangle\lyieldsw A\pf
\using\pcut_2
\endprooftree
\tb$\leadsto$\tb\\\\\\
\prooftree
\Gamma_1\lyieldsw\fbox{$P_1$}
\tb
\prooftree
\Delta|_k\vect{P_1}\lyields P_2\pf
\tb
\Gamma_2\langle\vect{P_2}\rangle\lyieldsw A
\justifies
\Gamma_2\langle\Delta|_k\vect{P_1}\rangle\lyieldsw A\pf
\using\ncut_1
\endprooftree
\justifies
\Gamma_2\langle\Delta|_k\Gamma_1\rangle\lyieldsw A\pf
\using\pcut_1
\endprooftree
}
The case of \sinfix{k}{} is entirely similar to the \scircum{k}{} case.\\
The case of \sinfix{k}{} is entirely similar to the \scircum{k}{} case.\\
\disp{ 
\prooftree
\prooftree
\Delta_1\lyieldsw\fbox{$P$}
\tb
\Delta_2\lyieldsw N
\justifies
\Delta_1|_k\Delta_2\lyieldsw\fbox{$P\swprod{k}N$}
\using\swprod{k}R
\endprooftree
\tb
\prooftree
\Gamma\langle\vect{P}|_k\vect{N}\rangle\lyieldsw A\pf
\justifies
\Gamma\langle\vect{P\swprod{k}N}\rangle\lyieldsw A\pf
\using\swprod{k}L
\endprooftree
\justifies
\Gamma\langle\Delta_1|_k\Delta_2\rangle\lyieldsw A\pf
\using\pcut_1
\endprooftree
\tb$\leadsto\tb$\\\\\\
\prooftree
\Delta_2\lyieldsw N
\tb
\prooftree
\Delta_1\lyieldsws \fbox{$P$}
\tb
\Gamma\langle\vect{P}|_k\vect{N}\rangle\lyieldsw A\pf
\justifies
\Gamma\langle\Delta_1|_k\vect{N}\rangle\lyieldsw A\pf
\using\pcut_1
\endprooftree
\justifies
\Gamma\langle\Delta_1|_k\Delta_2\rangle\lyieldsw A\pf
\using\ncut_2
\endprooftree
}\ \\
\disp{ 
$
\prooftree
\prooftree
\Delta\lyieldsw Q\pf
\tb
\Delta\lyieldsw A\pf
\justifies
\Delta\lyieldsw Q\aconj A\pf
\using\aconj R
\endprooftree
\tb
\prooftree
\Gamma\langle\vect{\fbox{$Q$}}\rangle\lyieldsw B
\justifies
\Gamma\langle\vect{\fbox{$Q\aconj A$}}\lyieldsw B
\using\aconj L
\endprooftree
\justifies
\Gamma\langle\Delta\rangle\lyieldsw B\pf
\using\pcut_2
\endprooftree
\tb\leadsto\tb\\\\\\
\prooftree
\Delta\lyields Q\pf\tb
\Gamma\langle\vect{\fbox{$Q$}}\rangle\lyieldsw B
\justifies
\Gamma\langle\Delta\rangle\lyieldsw B\pf
\using\pcut_2
\endprooftree
$
}
\disp{ 
$
\prooftree
\prooftree
\Delta\lyieldsw M\pf
\tb
\Delta\lyieldsw A\pf
\justifies
\Delta\lyieldsw M\aconj A\pf
\using\aconj R
\endprooftree
\tb
\prooftree
\Gamma\langle\vect M\rangle\lyieldsw B
\justifies
\Gamma\langle\vect{\fbox{$M\aconj A$}}\rangle\lyieldsw B
\using\aconj L
\endprooftree
\justifies
\Gamma\langle\Delta\rangle\lyieldsw B\pf
\using\pcut_2
\endprooftree
\tb\leadsto\tb\\\\\\
\prooftree
\Delta\lyields M\pf\tb
\Gamma\langle\vect M\rangle\lyieldsw B
\justifies
\Gamma\langle\Delta\rangle\lyieldsw B\pf
\using\ncut_1
\endprooftree
$
}
%
- Left commutative \pcut{} conversions:\\
\disp{
$
\prooftree
\prooftree
\Delta\langle\vect{\fbox{$Q$}}\rangle\lyieldsw N
\justifies
\Delta\langle\vect{Q}\rangle\lyieldsw N
\using\foc
\endprooftree
\tb 
\Gamma\langle\vect{\fbox{$N$}}\rangle\lyieldsw C
\justifies
\Gamma\langle\Delta\langle\vect{Q}\rangle\rangle\lyieldsw C
\using\pcut_2
\endprooftree
\tb\leadsto\tb\\\\\\
\prooftree
\prooftree
\Delta\langle\vect{\fbox{$Q$}}\rangle\lyieldsw N
\tb
\Gamma\langle\vect{\fbox{$N$}}\rangle\lyieldsw C
\justifies
\Gamma\langle\Delta\langle\vect{\fbox{$Q$}}\rangle\rangle\lyieldsw C
\using\pcut_2
\endprooftree
\justifies
\Gamma\langle\Delta\langle\vect{Q}\rangle\rangle\lyieldsw C
\using\foc
\endprooftree$
}
\disp{ 
$
\prooftree
\prooftree
\Delta\langle \vect{A}|_k\vect{B}\rangle\lyieldsw\fbox{$P$}
\justifies
\Delta\langle \vect{A\swprod{k}B}\rangle\lyieldsw\fbox{$P$}
\using\swprod{k}L
\endprooftree
\tb 
\Gamma\langle\vect{P}\rangle\lyieldsw C\pf
\justifies
\Gamma\langle\Delta\langle \vect{A\swprod{k}B}\rangle\rangle\lyieldsw C\pf
\using\pcut_1
\endprooftree
\tb\leadsto\tb\\\\\\
\prooftree
\prooftree
\Delta\langle\vect{A}|_k\vect{B}\rangle\lyieldsw\fbox{$P$}
\tb
\Gamma\langle\vect{P}\rangle\lyieldsw C\pf
\justifies
\Gamma\langle\Delta\langle \vect{A}|_k\vect{B}\rangle\rangle\lyieldsw C\pf
\using\pcut_1
\endprooftree
\justifies
\Gamma\langle\Delta\langle \vect{A\swprod{k}B}\rangle\rangle\lyieldsw C\pf
\using\swprod{k}L
\endprooftree
$
}
\disp{ 
$
\prooftree
\prooftree
\Delta\langle \vect{A}|_k\vect{B}\rangle\lyieldsw N\pf
\justifies
\Delta\langle \vect{A\swprod{k}B}\rangle\lyieldsw N\pf
\using\swprod{k}L
\endprooftree
\tb 
\Gamma\langle\fbox{$\vect{N}$}\rangle\lyieldsw C
\justifies
\Gamma\langle\Delta\langle \vect{A\swprod{k}B}\rangle\rangle\lyieldsw C\pf
\using\pcut_2
\endprooftree
\tb\leadsto\tb\\\\\\
\prooftree
\prooftree
\Delta\langle\vect{A}|_k\vect{B}\rangle\lyieldsw N\pf
\tb
\Gamma\langle\fbox{$\vect{N}$}\rangle\lyieldsw C
\justifies
\Gamma\langle\Delta\langle \vect{A}|_k\vect{B}\rangle\rangle\lyieldsw C\pf
\using\pcut_2
\endprooftree
\justifies
\Gamma\langle\Delta\langle \vect{A\swprod{k}B}\rangle\rangle\lyieldsw C\pf
\using\swprod{k}L
\endprooftree
$
}
\disp{ 
$
\prooftree
\prooftree
\Gamma_1\lyieldsw \fbox{$P_1$}
\tb
\Gamma_2\langle\vect{\fbox{$N_1$}}\rangle\lyieldsw N
\justifies
\Gamma_2\langle\vect{\fbox{$N_1\scircum{k}P_1$}}|_k\Gamma_1\rangle\lyieldsw N
\using\scircum{k}L
\endprooftree
\tb
\Theta\langle\vect{\fbox{$N$}}\rangle\lyieldsw C
\justifies
\Theta\langle\Gamma_2\langle\vect{\fbox{$N_1\scircum{k}P_1$}}|_k\Gamma_1\rangle\rangle\lyieldsw C
\using\pcut_2
\endprooftree
\tb\leadsto\tb\\\\\\
\prooftree
\Gamma_1\lyieldsw\fbox{$P_1$}
\tb
\prooftree
\Gamma_1\langle\vect{\fbox{$N_1$}}\rangle\lyieldsw N
\tb
\Theta\langle\vect{\fbox{$N$}}\rangle\lyieldsw C
\justifies
\Theta\langle\Gamma_2\langle\vect{\fbox{$N_1$}}\rangle\rangle\lyieldsw C
\using\pcut_2
\endprooftree
\justifies
\Theta\langle\Gamma_2\langle\vect{\fbox{$N_1\scircum{k}P_1$}}|_k\Gamma_1\rangle\rangle\lyieldsw C
\using\scircum{k}L
\endprooftree
$

}
\disp{ 
$
\prooftree
\prooftree
\Gamma\langle \vect A\rangle\lyieldsw \fbox{$P$}
\tb
\Gamma\langle \vect B\rangle\lyieldsw \fbox{$P$}
\justifies
\Gamma\langle \vect{A\adisj B}\rangle\lyieldsw \fbox{$P$}
\using\adisj L
\endprooftree
\tb
\Delta\langle\vect P\rangle\lyieldsw C\pf
\justifies
\Delta\langle\Gamma\langle \vect{A\adisj B}\rangle\rangle\lyieldsw C\pf
\using\pcut_1
\endprooftree
\tb\leadsto\tb\\\\\\
\prooftree
\prooftree
\Gamma\langle\vect A\rangle\lyieldsw\fbox{$P$}
\tb
\Delta\langle\vect P\rangle\lyieldsw C\pf
\justifies
\Delta\langle\Gamma\langle\vect A\rangle\rangle\lyieldsw C\pf
\using\pcut_1
\endprooftree
\tb
\prooftree
\Gamma\langle\vect B\rangle\lyieldsw\fbox{$P$}
\tb
\Delta\langle\vect P\rangle\lyieldsw C\pf
\justifies
\Delta\langle\Gamma\langle\vect B\rangle\rangle\lyieldsw C\pf
\using\pcut_1
\endprooftree
\justifies
\Delta\langle\Gamma\langle \vect{A\adisj B}\rangle\rangle\lyieldsw C\pf
\using\adisj L
\endprooftree
$
}

%
\disp{ 
$
\prooftree
\prooftree
\Gamma\langle \vect A\rangle\lyieldsw N\pf
\tb
\Gamma\langle \vect B\rangle\lyieldsw N\pf
\justifies
\Gamma\langle \vect{A\adisj B}\rangle\lyieldsw N\pf
\using\adisj L
\endprooftree
\tb
\Delta\langle\fbox{$\vect N$}\rangle\lyieldsw C
\justifies
\Delta\langle\Gamma\langle \vect{A\adisj B}\rangle\rangle\lyieldsw C\pf
\using\pcut_2
\endprooftree
\tb\leadsto\tb\\\\\\
\prooftree
\prooftree
\Gamma\langle\vect A\rangle\lyieldsw N\pf
\tb
\Delta\langle\fbox{$\vect N$}\rangle\lyieldsw C
\justifies
\Delta\langle\Gamma\langle\vect A\rangle\rangle\lyieldsw C\pf
\using\pcut_2
\endprooftree
\tb
\prooftree
\Gamma\langle\vect B\rangle\lyieldsw N\pf
\tb
\Delta\langle\fbox{$\vect N$}\rangle\lyieldsw C
\justifies
\Delta\langle\Gamma\langle\vect B\rangle\rangle\lyieldsw C\pf
\using\pcut_2
\endprooftree
\justifies
\Delta\langle\Gamma\langle \vect{A\adisj B}\rangle\rangle\lyieldsw C\pf
\using\adisj L
\endprooftree
$
}

- Right commutative \pcut{} conversions (unordered multiple distinguished occurrences are separated by semicolons):
\disp{ 
$
\prooftree
\Delta\lyieldsw\fbox{$P$}
\tb
\prooftree
\Gamma\langle\vect{P};\vect{\fbox{$Q$}}\rangle\lyieldsw C
\justifies
\Gamma\langle\vect{P};\vect{Q}\rangle\lyieldsw C
\using\foc
\endprooftree
\justifies
\Gamma\langle\Delta;\vect{Q}\rangle\lyieldsw C
\using\pcut_1
\endprooftree
\tb\leadsto\tb
\prooftree
\prooftree
\Delta\lyieldsw\fbox{$P$}
\tb
\Gamma\langle\vect{P};\vect{\fbox{$Q$}}\rangle\lyieldsw C
\justifies
\Gamma\langle\Delta;\vect{\fbox{$Q$}}\rangle\lyieldsw C
\using\pcut_1
\endprooftree
\justifies
\Gamma\langle\Delta;\vect{Q}\rangle\lyieldsw C
\using\foc
\endprooftree
$
}
\disp{ 
$
\prooftree
\Delta\lyieldsw\fbox{$P_1$}
\tb
\prooftree
\Gamma\langle\vect{P_1}\rangle\lyieldsw\fbox{$P_2$}
\justifies
\Gamma\langle\vect{P_1}\rangle\lyieldsw P_2
\using\foc
\endprooftree
\justifies
\Gamma\langle \Delta\rangle\lyieldsw P_2
\using\pcut_1
\endprooftree
\tb\leadsto\tb
\prooftree
\prooftree
\Delta\lyieldsw\fbox{$P_1$}
\tb
\Gamma\langle\vect{P_1}\rangle\lyieldsw \fbox{$P_2$}
\justifies
\Gamma\langle\Delta\rangle\lyieldsw \fbox{$P_2$}
\using\pcut_1
\endprooftree
\justifies
\Gamma\langle\Delta\rangle\lyieldsw P_2
\using\foc
\endprooftree
$
}
\disp{ 
$
\prooftree
\Delta\lyieldsw\fbox{$P$}
\tb
\prooftree
\Gamma\langle\vect{P}\rangle|_k\vect{A}\lyieldsw B\pf
\justifies
\Gamma\langle\vect{P}\rangle\lyieldsw B\scircum{k}A\pf
\using\scircum{k}R
\endprooftree
\justifies
\Gamma\langle\Delta\rangle\lyieldsw B\scircum{k}A\pf
\using\pcut_1
\endprooftree
\tb\leadsto\tb
\prooftree
\prooftree
\Delta\lyieldsw\fbox{$P$}
\tb
\Gamma\langle\vect{P}\rangle|_k \vect{A}\lyieldsw B\pf
\justifies
\Gamma\langle\Delta\rangle|_k \vect{A}\lyieldsw B\pf
\using\pcut_1
\endprooftree
\justifies
\Gamma\langle\Delta\rangle\lyieldsw B\scircum{k}A\pf
\using\scircum{k}R
\endprooftree
$
}
\disp{ 
$
\prooftree
\Delta\lyieldsw N\pf
\tb
\prooftree
\Gamma\langle\fbox{$\vect{N}$}\rangle|_k\vect{A}\lyieldsw B
\justifies
\Gamma\langle\fbox{$\vect{N}$}\rangle\lyieldsw B\scircum{k}A
\using\scircum{k}R
\endprooftree
\justifies
\Gamma\langle\Delta\rangle\lyieldsw B\scircum{k}A\pf
\using\pcut_2
\endprooftree
\tb\leadsto\tb
\prooftree
\prooftree
\Delta\lyieldsw N\pf
\tb
\Gamma\langle\fbox{$\vect{N}$}\rangle|_k \vect{A}\lyieldsw B
\justifies
\Gamma\langle\Delta\rangle|_k \vect{A}\lyieldsw B\pf
\using\pcut_2
\endprooftree
\justifies
\Gamma\langle\Delta\rangle\lyieldsw B\scircum{k}A\pf
\using\scircum{k}R
\endprooftree
$
}
\disp{ 
$
\prooftree
\Delta\lyieldsw\fbox{$P$}
\tb
\prooftree
\Gamma\langle\vect{P};\vect{A}|_k\vect{B}\rangle\lyieldsw C\pf
\justifies
\Gamma\langle\vect{P};\vect{A\swprod{k}B}\rangle\lyieldsw C\pf
\using\swprod{k}L
\endprooftree
\justifies
\Gamma\langle\Delta;\vect{A\swprod{k}B}\rangle\lyieldsw C\pf
\using\pcut_1
\endprooftree
\tb\leadsto\tb
\prooftree
\prooftree
\Delta\lyieldsw\fbox{$P$}
\tb
\Gamma\langle\vect{P};\vect{A}|_k\vect{B}\rangle\lyieldsw C\pf
\justifies
\Gamma\langle\Delta;\vect{A}|_k\vect{B}\rangle\lyieldsw C\pf
\using\pcut_1
\endprooftree
\justifies
\Gamma\langle\Delta;\vect{A\swprod{k}B}\rangle\lyieldsw C\pf
\using\swprod{k}L
\endprooftree
$
}
\disp{
$
\prooftree
\Delta\lyieldsw N\pf
\tb
\prooftree
\Gamma\langle\vect{\fbox{$N$}};\vect{A}|_k\vect{B}\rangle\lyieldsw C
\justifies
\Gamma\langle\vect{\fbox{$N$}};\vect{A\swprod{k}B}\rangle\lyieldsw C
\using\swprod{k}L
\endprooftree
\justifies
\Gamma\langle\Delta;\vect{A\swprod{k}B}\rangle\lyieldsw C\pf
\using\pcut_2
\endprooftree
\tb\leadsto\tb
\prooftree
\prooftree
\Delta\lyieldsw N\pf\tb\Gamma\langle\vect{\fbox{$N$}};\vect{A}|_k\vect{B}\rangle\lyieldsw C
\justifies
\Gamma\langle\Delta;\vect{A}|_k\vect{B}\rangle\lyieldsw C\pf
\using\pcut_2
\endprooftree
\justifies
\Gamma\langle\Delta;\vect{A\swprod{k}B}\rangle\lyieldsw C\pf
\using\swprod{k}L
\endprooftree
$
}
\disp{
$
\prooftree
\Delta\lyieldsw\fbox{$P$}
\tb
\prooftree
\Gamma\lyieldsw\fbox{$P_1$}
\tb
\Theta\langle\vect{P_2};\vect{P}\rangle\lyieldsw C
\justifies
\Theta\langle\vect{\fbox{$P_2\scircum{k}P_1$}}|_k\Gamma;\vect{P}\rangle\lyieldsw C
\using\scircum{k}L
\endprooftree
\justifies
\Theta\langle\vect{\fbox{$P_2\scircum{k}P_1$}}|_k\Gamma;\Delta\rangle\lyieldsw C
\using\pcut_1
\endprooftree
\tb\leadsto\tb\\\\\\
\prooftree
\Gamma\lyieldsw\fbox{$P_1$}
\tb
\prooftree
\Delta\lyieldsw\fbox{$P$}
\tb
\Theta\langle\vect{P_2};\vect{P}\rangle\lyieldsw C
\justifies
\Theta\langle\vect{P_2};\Delta\rangle\lyieldsw C
\using\pcut_1
\endprooftree
\justifies
\Theta\langle\vect{\fbox{$P_2\scircum{k}P_1$}}|_k\Gamma;\Delta\rangle\lyieldsw C
\using\scircum{k}L
\endprooftree
$
}
\disp{ 
$
\prooftree
\Delta\lyieldsw\fbox{$P$}
\tb
\prooftree
\Gamma\langle\vect P\rangle\lyieldsw A\pf
\tb
\Gamma\langle\vect P\rangle\lyieldsw B\pf
\justifies
\Gamma\langle\vect P\rangle\lyieldsw A\aconj B\pf
\using\aconj R
\endprooftree
\justifies
\Gamma\langle\Delta\rangle\lyieldsw A\aconj B\pf
\using\pcut_1
\endprooftree
\tb\leadsto\tb\\\\\\
\prooftree
\prooftree
\Delta\lyieldsw \fbox{$P$}
\tb
\Gamma\langle\vect P\rangle\lyieldsw A\pf
\justifies
\Gamma\langle\Delta\rangle\lyieldsw A\pf
\using\pcut_1
\endprooftree
\tb
\prooftree
\Delta\lyieldsw \fbox{$P$}
\tb
\Gamma\langle\vect P\rangle\lyieldsw B\pf
\justifies
\Gamma\langle\Delta\rangle\lyieldsw B\pf
\using\pcut_1
\endprooftree
\justifies
\Gamma\langle\Delta\rangle\lyieldsw A\aconj B\pf
\using\aconj R
\endprooftree
$
}
\disp{ 
$
\prooftree
\Delta\lyieldsw N\pf
\tb
\prooftree
\Gamma\langle\fbox{$\vect N$}\rangle\lyieldsw A
\tb
\Gamma\langle\fbox{$\vect N$}\rangle\lyieldsw B
\justifies
\Gamma\langle\fbox{$\vect N$}\rangle\lyieldsw A\aconj B
\using\aconj R
\endprooftree
\justifies
\Gamma\langle\Delta\rangle\lyieldsw A\aconj B\pf
\using\pcut_2
\endprooftree
\tb\leadsto\tb\\\\\\
\prooftree
\prooftree
\Delta\lyieldsw N\pf
\tb
\Gamma\langle\fbox{$\vect N$}\rangle\lyieldsw A
\justifies
\Gamma\langle\Delta\rangle\lyieldsw A\pf
\using\pcut_2
\endprooftree
\tb
\prooftree
\Delta\lyieldsw N\pf
\tb
\Gamma\langle\fbox{$\vect N$}\rangle\lyieldsw B
\justifies
\Gamma\langle\Delta\rangle\lyieldsw B\pf
\using\pcut_2
\endprooftree
\justifies
\Gamma\langle\Delta\rangle\lyieldsw A\aconj B\pf
\using\aconj R
\endprooftree
$
}
- Left commutative \ncut{} conversions:\\
\disp{
$
\prooftree
\prooftree
\Delta\langle\vect{\fbox{$Q$}}\rangle\lyieldsw P
\justifies
\Delta\langle\vect{Q}\rangle\lyieldsw P
\using\foc
\endprooftree
\tb
\Gamma\langle\vect{P}\rangle\lyieldsw C
\justifies
\Gamma\langle\Delta\langle \vect{Q}\rangle\rangle\lyieldsw C
\using\ncut_1
\endprooftree
\tb\leadsto\tb
\prooftree
\prooftree
\Delta\vect{\fbox{$Q$}}\lyieldsw P
\tb
\Gamma\langle\vect{P}\rangle\lyieldsw C
\justifies
\Gamma\langle\Delta\langle \vect{\fbox{$Q$}}\rangle\rangle\lyieldsw C
\using\ncut_1
\endprooftree
\justifies
\Gamma\langle\Delta\langle \vect{Q}\rangle\rangle\lyieldsw C
\using\foc
\endprooftree
$
}
\disp{ 
$
\prooftree
\prooftree
\Delta\langle \vect{A}|_k\vect{B}\rangle\lyieldsw P\pf
\justifies
\Delta\langle \vect{A\swprod{k}B}\rangle\lyieldsw P\pf
\using\swprod{k}L
\endprooftree
\tb 
\Gamma\langle\vect{P}\rangle\lyieldsw C
\justifies
\Gamma\langle\Delta\langle \vect{A\swprod{k}B}\rangle\rangle\lyieldsw C\pf
\using\ncut_1
\endprooftree
\tb\leadsto\tb\\\\\\
\prooftree
\prooftree
\Delta\langle\vect{A}|_k\vect{B}\rangle\lyieldsw P\pf
\tb
\Gamma\langle\vect{P}\rangle\lyieldsw C
\justifies
\Gamma\langle\Delta\langle \vect{A}|_k\vect{B}\rangle\rangle\lyieldsw C\pf
\using\ncut_1
\endprooftree
\justifies
\Gamma\langle\Delta\langle \vect{A\swprod{k}B}\rangle\rangle\lyieldsw C\pf
\using\swprod{k}L
\endprooftree
$
}
\disp{ 
$
\prooftree
\prooftree
\Delta\langle \vect{A}|_k\vect{B}\rangle\lyieldsw N
\justifies
\Delta\langle \vect{A\swprod{k}B}\rangle\lyieldsw N
\using\swprod{k}L
\endprooftree
\tb 
\Gamma\langle\vect{N}\rangle\lyieldsw C\pf
\justifies
\Gamma\langle\Delta\langle \vect{A\swprod{k}B}\rangle\rangle\lyieldsw C\pf
\using\ncut_2
\endprooftree
\tb\leadsto\tb\\\\\\
\prooftree
\prooftree
\Delta\langle\vect{A}|_k\vect{B}\rangle\lyieldsw N
\tb
\Gamma\langle\vect{N}\rangle\lyieldsw C\pf
\justifies
\Gamma\langle\Delta\langle \vect{A}|_k\vect{B}\rangle\rangle\lyieldsw C\pf
\using\ncut_2
\endprooftree
\justifies
\Gamma\langle\Delta\langle \vect{A\swprod{k}B}\rangle\rangle\lyieldsw C\pf
\using\swprod{k}L
\endprooftree
$
}
\disp{ 
$
\prooftree
\prooftree
\Gamma_1\lyieldsw \fbox{$P_1$}
\tb
\Gamma_2\langle\vect{\fbox{$N_1$}}\rangle\lyieldsw P
\justifies
\Gamma_2\langle\vect{\fbox{$N_1\scircum{k}P_1$}}|_k\Gamma_1\rangle\lyieldsw P
\using\scircum{k}L
\endprooftree
\tb
\Theta\langle\vect{P}\rangle\lyieldsw C
\justifies
\Theta\langle\Gamma_2\langle\vect{\fbox{$N_1\scircum{k}P_1$}}|_k\Gamma_1\rangle\rangle\lyieldsw C
\using\ncut_1
\endprooftree
\tb\leadsto\tb\\\\\\
\prooftree
\Gamma_1\lyieldsw\fbox{$P_1$}
\tb
\prooftree
\Gamma_1\langle\vect{\fbox{$N_1$}}\rangle\lyieldsw P
\tb
\Theta\langle\vect{P}\rangle\lyieldsw C
\justifies
\Theta\langle\Gamma_2\langle\vect{\fbox{$N_1$}}\rangle\rangle\lyieldsw C
\using\ncut_1
\endprooftree
\justifies
\Theta\langle\Gamma_2\langle\vect{\fbox{$N_1\scircum{k}P_1$}}|_k\Gamma_1\rangle\rangle\lyieldsw C
\using\scircum{k}L
\endprooftree
$
}
%
\disp{ 
$
\prooftree
\prooftree
\Gamma\langle \vect A\rangle\lyieldsw P\pf
\tb
\Gamma\langle \vect B\rangle\lyieldsw P\pf
\justifies
\Gamma\langle \vect{A\adisj B}\rangle\lyieldsw P\pf
\using\adisj L
\endprooftree
\tb
\Delta\langle\vect P\rangle\lyieldsw C\pf
\justifies
\Delta\langle\Gamma\langle \vect{A\adisj B}\rangle\rangle\lyieldsw C\pf
\using\ncut_1
\endprooftree
\tb\leadsto\tb\\\\\\
\prooftree
\prooftree
\Gamma\langle\vect A\rangle\lyieldsw P\pf
\tb
\Delta\langle\vect P\rangle\lyieldsw C
\justifies
\Delta\langle\Gamma\langle\vect A\rangle\rangle\lyieldsw C\pf
\using\ncut_1
\endprooftree
\tb
\prooftree
\Gamma\langle\vect B\rangle\lyieldsw P\pf
\tb
\Delta\langle\vect P\rangle\lyieldsw C
\justifies
\Delta\langle\Gamma\langle\vect B\rangle\rangle\lyieldsw C\pf
\using\ncut_1
\endprooftree
\justifies
\Delta\langle\Gamma\langle \vect{A\adisj B}\rangle\rangle\lyieldsw C\pf
\using\adisj L
\endprooftree
$
}

%
\disp{ 
$
\prooftree
\prooftree
\Gamma\langle \vect A\rangle\lyieldsw N
\tb
\Gamma\langle \vect B\rangle\lyieldsw N
\justifies
\Gamma\langle \vect{A\adisj B}\rangle\lyieldsw N
\using\adisj L
\endprooftree
\tb
\Delta\langle \vect N\rangle\lyieldsw C\pf
\justifies
\Delta\langle\Gamma\langle \vect{A\adisj B}\rangle\rangle\lyieldsw C\pf
\using\ncut_2
\endprooftree
\tb\leadsto\tb\\\\\\
\prooftree
\prooftree
\Gamma\langle\vect A\rangle\lyieldsw N
\tb
\Delta\langle\vect N\rangle\lyieldsw C\pf
\justifies
\Delta\langle\Gamma\langle\vect A\rangle\rangle\lyieldsw C\pf
\using\ncut_2
\endprooftree
\tb
\prooftree
\Gamma\langle\vect B\rangle\lyieldsw N
\tb
\Delta\langle\vect N\rangle\lyieldsw C\pf
\justifies
\Delta\langle\Gamma\langle\vect B\rangle\rangle\lyieldsw C\pf
\using\ncut_2
\endprooftree
\justifies
\Delta\langle\Gamma\langle \vect{A\adisj B}\rangle\rangle\lyieldsw C\pf
\using\adisj L
\endprooftree
$
}

- Right commutative \ncut{} conversions:\\
\disp{ 
$
\prooftree
\Delta\lyieldsw N
\tb
\prooftree
\Gamma\langle\vect{N};\vect{\fbox{$Q$}}\rangle\lyieldsw C
\justifies
\Gamma\langle\vect{N};\vect{Q}\rangle\lyieldsw C
\using\foc
\endprooftree
\justifies
\Gamma\langle\Delta;\vect{Q}\rangle\lyieldsw C
\using\ncut_2
\endprooftree
\tb\leadsto\tb
\prooftree
\prooftree
\Delta\lyieldsw N
\tb
\Gamma\langle\vect{N};\vect{\fbox{$Q$}}\rangle\lyieldsw C
\justifies
\Gamma\langle\Delta;\vect{\fbox{$Q$}}\rangle\lyieldsw C
\using\ncut_2
\endprooftree
\justifies
\Gamma\langle\Delta;\vect{Q}\rangle\lyieldsw C
\using\foc
\endprooftree
$
}
\disp{ 
$
\prooftree
\Delta\lyieldsw N
\tb
\prooftree
\Gamma\langle\vect{N}\rangle\lyieldsw\fbox{$P$}
\justifies
\Gamma\langle\vect{N}\rangle\lyieldsw P
\using\foc
\endprooftree
\justifies
\Gamma\langle\Delta\rangle\lyieldsw P
\using\ncut_2
\endprooftree
\tb\leadsto\tb
\prooftree
\prooftree
\Delta\lyieldsw N
\tb
\Gamma\langle\vect{N}\rangle\lyieldsw \fbox{$P$}
\justifies
\Gamma\langle\Delta\rangle\lyieldsw \fbox{$P$}
\using\ncut_2
\endprooftree
\justifies
\Gamma\langle\Delta\rangle\lyieldsw P
\using\foc
\endprooftree
$
}
\disp{ 
$
\prooftree
\Delta\lyieldsw P\pf
\tb
\prooftree
\Gamma\langle\vect{P}\rangle|_k\vect{A}\lyieldsw B
\justifies
\Gamma\langle\vect{P}\rangle\lyieldsw B\scircum{k}A
\using\scircum{k}R
\endprooftree
\justifies
\Gamma\langle\Delta\rangle\lyieldsw B\scircum{k}A\pf
\using\ncut_1
\endprooftree
\tb\leadsto\tb
\prooftree
\prooftree
\Delta\lyieldsw P\pf
\tb
\Gamma\langle\vect{P}\rangle|_k \vect{A}\lyieldsw B
\justifies
\Gamma\langle\Delta\rangle|_k \vect{A}\lyieldsw B\pf
\using\ncut_1
\endprooftree
\justifies
\Gamma\langle\Delta\rangle\lyieldsw B\scircum{k}A\pf
\using\scircum{k}R
\endprooftree
$
}
\disp{ 
$
\prooftree
\Delta\lyieldsw N
\tb
\prooftree
\Gamma\langle\vect{P}\rangle|_k\vect{A}\lyieldsw B\pf
\justifies
\Gamma\langle\vect{P}\rangle\lyieldsw B\scircum{k}A\pf
\using\scircum{k}R
\endprooftree
\justifies
\Gamma\langle\Delta\rangle\lyieldsw B\scircum{k}A\pf
\using\ncut_2
\endprooftree
\tb\leadsto\tb
\prooftree
\prooftree
\Delta\lyieldsw N
\tb
\Gamma\langle\vect{P}\rangle|_k \vect{A}\lyieldsw B\pf
\justifies
\Gamma\langle\Delta\rangle|_k \vect{A}\lyieldsw B\pf
\using\ncut_2
\endprooftree
\justifies
\Gamma\langle\Delta\rangle\lyieldsw B\scircum{k}A\pf
\using\scircum{k}R
\endprooftree
$
}
\disp{ 
$
\prooftree
\Delta\lyieldsw P\pf
\tb
\prooftree
\Gamma\langle\vect{P};\vect{A}|_k\vect{B}\rangle\lyieldsw C
\justifies
\Gamma\langle\vect{P};\vect{A\swprod{k}B}\rangle\lyieldsw C
\using\swprod{k}L
\endprooftree
\justifies
\Gamma\langle\Delta;\vect{A\swprod{k}B}\rangle\lyieldsw C\pf
\using\ncut_1
\endprooftree
\tb\leadsto\tb
\prooftree
\prooftree
\Delta\lyieldsw P\pf
\tb
\Gamma\langle\vect{P};\vect{A}|_k\vect{B}\rangle\lyieldsw C
\justifies
\Gamma\langle\Delta;\vect{A}|_k\vect{B}\rangle\lyieldsw C\pf
\using\ncut_1
\endprooftree
\justifies
\Gamma\langle\Delta;\vect{A\swprod{k}B}\rangle\lyieldsw C\pf
\using\swprod{k}L
\endprooftree
$
}
\disp{ 
$
\prooftree
\Delta\lyieldsw N
\tb
\prooftree
\Gamma\langle\vect{N};\vect{A}|_k\vect{B}\rangle\lyieldsw C\pf
\justifies
\Gamma\langle\vect{N};\vect{A\swprod{k}B}\rangle\lyieldsw C\pf
\using\swprod{k}L
\endprooftree
\justifies
\Gamma\langle\Delta;\vect{A\swprod{k}B}\rangle\lyieldsw C\pf
\using\ncut_2
\endprooftree
\tb\leadsto\tb
\prooftree
\prooftree
\Delta\lyieldsw N\tb\Gamma\langle\vect{N};\vect{A}|_k\vect{B}\rangle\lyieldsw C\pf
\justifies
\Gamma\langle\Delta;\vect{A}|_k\vect{B}\rangle\lyieldsw C\pf
\using\ncut_2
\endprooftree
\justifies
\Gamma\langle\Delta;\vect{A\swprod{k}B}\rangle\lyieldsw C\pf
\using\swprod{k}L
\endprooftree
$
}
\disp{
$
\prooftree
\Delta\lyieldsw N
\tb
\prooftree
\Gamma\lyieldsw\fbox{$P_1$}
\tb
\Theta\langle\vect{P_2};\vect{N}\rangle\lyieldsw C
\justifies
\Theta\langle\vect{\fbox{$P_2\scircum{k}P_1$}}|_k\Gamma;\vect{N}\rangle\lyieldsw C
\using\scircum{k}L
\endprooftree
\justifies
\Theta\langle\vect{\fbox{$P_2\scircum{k}P_1$}}|_k\Gamma;\Delta\rangle\lyieldsw C
\using\ncut_2
\endprooftree
\tb\leadsto\tb\\\\\\
\prooftree
\Gamma\lyieldsw\fbox{$P_1$}
\tb
\prooftree
\Delta\lyieldsw N
\tb
\Theta\langle\vect{P_2};\vect{N}\rangle\lyieldsw C
\justifies
\Theta\langle\vect{P_2};\Delta\rangle\lyieldsw C
\using\ncut_2
\endprooftree
\justifies
\Theta\langle\vect{\fbox{$P_2\scircum{k}P_1$}}|_k\Gamma;\Delta\rangle\lyieldsw C
\using\scircum{k}L
\endprooftree
$}
\disp{ 
$
\prooftree
\Delta\lyieldsw P\pf
\tb
\prooftree
\Gamma\langle\vect P\rangle\lyieldsw A
\tb
\Gamma\langle\vect P\rangle\lyieldsw B
\justifies
\Gamma\langle\vect P\rangle\lyieldsw A\aconj B
\using\aconj R
\endprooftree
\justifies
\Gamma\langle\Delta\rangle\lyieldsw A\aconj B\pf
\using\ncut_1
\endprooftree
\tb\leadsto\tb\\\\\\
\prooftree
\prooftree
\Delta\lyieldsw P
\tb
\Gamma\langle\vect P\rangle\lyieldsw A
\justifies
\Gamma\langle\Delta\rangle\lyieldsw A\pf
\using\ncut_1
\endprooftree
\tb
\prooftree
\Delta\lyieldsw P\pf
\tb
\Gamma\langle\vect P\rangle\lyieldsw B
\justifies
\Gamma\langle\Delta\rangle\lyieldsw B\pf
\using\ncut_1
\endprooftree
\justifies
\Gamma\langle\Delta\rangle\lyieldsw A\aconj B\pf
\using\aconj R
\endprooftree
$
}
\disp{ 
$
\prooftree
\Delta\lyieldsw N
\tb
\prooftree
\Gamma\langle\vect N\rangle\lyieldsw A\pf
\tb
\Gamma\langle\vect N\rangle\lyieldsw B\pf
\justifies
\Gamma\langle\vect N\rangle\lyieldsw A\aconj B\pf
\using\aconj R
\endprooftree
\justifies
\Gamma\langle\Delta\rangle\lyieldsw A\aconj B\pf
\using\ncut_2
\endprooftree
\tb\leadsto\tb\\\\\\
\prooftree
\prooftree
\Delta\lyieldsw N
\tb
\Gamma\langle\vect N\rangle\lyieldsw A\pf
\justifies
\Gamma\langle\Delta\rangle\lyieldsw A\pf
\using\ncut_2
\endprooftree
\tb
\prooftree
\Delta\lyieldsw N
\tb
\Gamma\langle\vect N\rangle\lyieldsw B\pf
\justifies
\Gamma\langle\Delta\rangle\lyieldsw B\pf
\using\ncut_2
\endprooftree
\justifies
\Gamma\langle\Delta\rangle\lyieldsw A\aconj B\pf
\using\aconj R
\endprooftree
$
}\ \\
This completes the proof.}

\end{document}